# CRYSTAL BASES, DILOGARITHM IDENTITIES AND TORSION IN ALGEBRAIC *K*–GROUPS

EDWARD FRENKEL AND ANDRÁS SZENES

April, 1993

## 1. INTRODUCTION

In this paper we give a new interpretation and proof of the dilogarithm identities

$$(1.1) \qquad \sum_{j=1}^{k} L(\delta_j) = \frac{\pi^2}{6} \frac{3k}{k+2},$$

where $L(z)$ is the Rogers dilogarithm function:

$$L(z) = \int_0^z \log w \, d\log(1-w) - \frac{1}{2} \log z \log(1-z), \quad 0 \leq z \leq 1,$$

and $\delta_j = \sin^2 \frac{\pi}{k+2} / \sin^2 \frac{\pi(j+1)}{k+2}$. They were first proved by Kirillov and Reshetikhin [1].

The right hand side of this identity is equal to $\frac{\pi^2}{6}$ times the central charge of an integrable level $k$ representation of the affine Kac-Moody Lie algebra $\hat{\mathfrak{sl}}_2$. This number appears in the asymptotics of the character of such a representation. Following the general approach taken in our previous work [2], we will obtain the left hand side as a result of an alternative calculation of this asymptotics.

We will use a combinatorial description of the so-called crystal basis [6] of an integrable $\hat{\mathfrak{sl}}_2$-module $V^k$ of level $k$, found by Jimbo, Misra, Miwa and Okado in [3] (cf. also [4, 5]). They defined a weight function $\omega$ on the crystal basis vectors, such that if we sum up $q^\omega$ over all of them, we obtain the character of $V^k$. We will introduce an increasing system of finite subsets of the set of all crystal basis vectors and show that the corresponding partial characters are related by a $q$-recurrence relation. This will give us a formula for the character of $V^k$ via the product of infinitely many matrices. In the limit when $q \to 1$, we will obtain a formula for the asymptotics as an integral, which is then shown to be equal to the left hand side of (1.1).

The dilogarithm identities are closely connected with elements of torsion in the group $K_3(F)$, where $F$ is a totally real field of algebraic numbers. We construct such

Research of the first author was supported by a Junior Fellowship from the Harvard Society of Fellows and by NSF grant DMS-9205303

E-mail addresses: frenkel@math.harvard.edu, szenes@math.mit.edu

elements explicitly using the so-called Bloch group and the relative group $K_2$ of the projective line over $F$ modulo two points. We can then interpret the left hand side of the identity (1.1) and our integral as the images of these elements under certain homomorphisms from $K_3(F)$ to $\mathbb{R}/(\mathbb{Z}\frac{\pi^2}{6})$.

The paper is organized as follows: Section 2 contains all necessary facts about the path description of the crystal basis of $V^k$ as defined in [3], and the definition of our subsets. We use them to obtain a new character formula via the product of infinitely many matrices. In Section 3 we derive from this formula an integral representation of the asymptotics of the character when $q \to 1$. In Section 4 we evaluate this integral as a sum of dilogarithms. In Section 5 we discuss the connection between the dilogarithm identities and torsion in the group $K_3$ of totally real number fields. Finally, in Section 6, we discuss generalizations of our results to higher rank affine Lie algebras. The Appendix contains the most technical part of the proof of the main result of Section 3.

## 2. Path realizations of crystal bases

Let $k$ be a positive integer. Denote by $V^k$ the vacuum representation (representation with highest weight 0) of the affine Kac-Moody Lie algeba $\hat{\mathfrak{sl}}_2$ (cf. [7]). This is an infinite-dimensional $\mathbb{Z}$-graded vector space $V^k = \oplus_{n=0}^{\infty} V^k(n)$. We can define its character by $\operatorname{ch} V^k = \sum_{n=0}^{\infty} \dim V^k(n) q^n$. Note that this definition can be applied to any $\mathbb{Z}$-graded vector space.

In [3] an explicit description of the crystal basis of $V^k$ is given. Define $S^k = \{0, 1, \ldots, k\}$, the set of integers from 0 to $k$. A path is a sequence $\eta = (\eta_0, \eta_1, \ldots)$, where $\eta_m \in S^k$. Let us introduce the path $\mu = (k, 0, k, 0, \ldots)$. A path $\eta$ is called restricted, if for some $N$, $\eta_m = \mu_m$, when $m > N$. Let $R^k$ be the set of all restricted paths. Define the weight function $\omega$ on $R^k$ by the formula

$$(2.1) \qquad \omega(\eta) = \sum_{m=1}^{\infty} m[H(\eta_{m-1}, \eta_m) - H(\mu_{m-1}, \mu_m)],$$

where $H(\nu, \nu') = \max(k - \nu, \nu')$.

It was shown in [3] (cf. also [4]) that the restricted paths are in one-to-one correspondence with the elements of the crystal basis of $V^k$. Moreover, the degree of the basis vector corresponding to a path $\eta$ is given by the weight function $\omega$ defined above. This implies the following result:

**Theorem 2.1** ([3]). *As formal power series in $q$*

$$\operatorname{ch} V^k = \sum_{\eta \in R^k} q^{\omega(\eta)}.$$

Now we introduce the subsets $W_N^r \in R^k$, $N = 1, \ldots$ and $r \in S^k$. The subset $W_N^r$ consists of paths $\eta$ such that $\eta_j = \mu_j$ for $j \geq 2N$ and $\eta_{2N-1} = r$. Denote by $w_N^r$ the character of $W_N^r$ and let $\mathbf{w}_N$ be the column vector $[w_N^0, w_N^1, \ldots, w_N^k]$.

**Lemma 2.2.** *We have the following recursion:*

$$\mathbf{w}_{N+1} = q^{-2kN} M_k(q^{2N+1}) M_k(q^{2N}) \mathbf{w}_N,$$

*where the $(i,j)$th entry of the matrix $M_k(x)$ is given by*

$$M_k(x)_{i,j} = x^{H(j,i)}, \quad i,j = 0, 1, \ldots, k.$$

*Proof.* The Lemma follows from the definition of the weight function (2.1). The multiplier $q^{-2kN}$ enters, since $H(\mu_{2N-1}, \mu_{2N}) = k$ and $H(\mu_{2N}, \mu_{2N+1}) = 0$. □

Introduce the notation $L_k(q, x) = x^{-2k} M_k(qx^2) M_k(x^2)$ and the starting vector $\mathbf{w}_0 = [1, 0, \ldots, 0]$. The inductive formula of the Lemma gives us a new character formula:

(2.2) $$\operatorname{ch} V^k = \mathbf{w}_0^t \ldots L_k(q, q^N) \ldots L_k(q, q) L_k(q, 1) \mathbf{w}_0.$$

*Remark* 2.1. It is known that the character of $V^k$ is an analytic function of $q$ for $|q| < 1$. In fact, up to a factor of a power of $q$, it is equal to a modular function in $\tau$, where $q = e^{2\pi i \tau}$, with respect to a congruence subgroup [7]. For instance, the character of $V^1$,

$$\operatorname{ch} V^1 = q^{1/24} \frac{\theta(\tau)}{\eta(\tau)},$$

is the ratio of the theta-function and the Dedekind's eta-function [7].

Thus formula (2.2) gives a new matrix product expression for these modular functions. It would be interesting to know, under what conditions such expressions exists in general. □

Using the modular properties of $\operatorname{ch} V^k$, Kac and Wakimoto proved the following result.

**Theorem 2.3** ([8]).

$$\lim_{q \to 1} (1-q) \log \operatorname{ch} V^k = \frac{\pi^2}{6} \frac{3k}{k+2}.$$

The number $3k/(k+2)$ in the right hand side of this formula is equal to the central charge of the Virasoro algebra, which acts on $V^k$ through the Sugawara construction.

In the next two Sections we will derive a new formula for this asymptotics.

## 3. Asymptotics of infinite product of matrices

The following proposition relates the asymptotics of the infinite product (2.2) to the asymptotics of the product of the highest eigenvalues of $L_k(1,x) = x^{-2k}M_k(x^2)^2$.

**Proposition 3.1.** *Denote by $d_k(x)$ the eigenvalue of $L_k(1,x)$ of maximal absolute value. Then $d_k(x)$ is a positive analytic function of $x$, for $0 \leq x \leq 1$. The asymptotic behavior of $\operatorname{ch} V^k$ is the same as that of the infinite product of the $d_k(q^N)$, more precisely*

$$\lim_{q \to 1}(1-q)\log \operatorname{ch} V^k = \lim_{q \to 1}(1-q)\log \prod_{N=0}^{\infty} d_k(q^N).$$

*Remark* 3.1. We can write down an integral representation of the asymptotics using the Proposition as follows: the expression on the right hand side under the limit can be replaced by

$$-\log q \sum_{N=0}^{\infty} \log d_k(q^N).$$

In this form, this is manifestly a Riemann sum of the form $\log d_k(x)\, d\log x$. Thus, taking the limit $q \to 1$, we obtain

(3.1) $$\lim_{q \to 1}(1-q)\log \operatorname{ch} V^k = \int_{x=0}^{x=1} \log d_k(x)\, d\log x.$$

□

The proof of Proposition 3.1 is quite technical. We will give it in two parts. In the main text, we include two key lemmas, which immediately imply the analog of Proposition 3.1 for a family of matrices $B(x)$ satisfying some extra conditions. The proof of the proposition will be completed in the Appendix.

Denote by $D_k$ the set of real $(k+1) \times (k+1)$ matrices, which have $k+1$ different real, positive eigenvalues. Assume that $B : [0,1] \to D_k$ is a smooth family of such matrices. Denote the normalized eigenvectors of $B(x)$ by $\mathbf{e}^i(x)$, $i = 0, 1, \ldots, k$, with corresponding eigenvalues $d^i(x)$. Because of the assumption on $B(x)$, we can assume that $d^0(x) > d^1(x) > \cdots > d^k(x) > 0$. Then $d^i$ and $\mathbf{e}^i$ depend smoothly on $x$. Fix a value of $q$ between 0 and 1, and introduce a starting vector $\mathbf{w}_0$ close to $\mathbf{e}^0(1)$, and $\mathbf{w}_N = B(q^{N-1})\ldots B(q)B(1)\mathbf{w}_0$. Our goal is to compare

$$\lim_{N \to \infty} \log \mathbf{e}^0(0)^t \mathbf{w}_N \quad \text{and} \quad \log \prod_{N=0}^{\infty} d^0(q^N).$$

The idea is to show that $\mathbf{w}_N/|\mathbf{w}_N|$ is close to $\mathbf{e}^0(q^N)$, and then to argue that $\mathbf{w}_{N+1} \sim d^0(q^N)\mathbf{w}_N$. To formalize this, introduce the matrix $A(x)$ with column vectors $\mathbf{e}^i(x)$. Then $D(x) = A^{-1}(x)B(x)A(x)$ is a diagonal matrix with the eigenvalues $d^0(x),\ldots,d^k(x)$ on the diagonal. Let $\mathbf{u}_N = A(q^N)^{-1}\mathbf{w}_N$, and denote by $u_N^i$, $i =$ $0, \ldots, k$, its components, i.e. the coordinates of $\mathbf{w}_N$ in the basis of eigenvectors of $B(q^N)$.

Introduce the following important quantities:
$$\alpha = \max_{x \in [0,1], 0 < i \leq k} \frac{d^i(x)}{d^0(x)}, \quad \text{and} \quad \beta = \max_{x \in [0,1]} \left| \frac{d}{dx} \frac{1}{\det(A(x))} \right|,$$

describing the family $B(x)$. We will also use a series of constants $C_*$, each of which can be bounded by a universal (independent of $B$) polynomial of $m_B = \max_{x \in [0,1]} \{|B_{ij}(x)|, 0 \leq i, j \leq k\}$. Note that from now on, we always assume that $q$ is sufficiently close to 1.

**Lemma 3.2.** *There is a constant $C_\mathrm{r}$, such that if $\gamma$ satisfies $\gamma|u_0^0| > |u_0^i|$ for $i > 1$, and*

(3.2) $$C_\mathrm{r}(1-q) < \gamma \frac{1-\alpha}{\beta},$$

*then*
$$\left| \frac{u_N^i}{u_N^0} \right| < \gamma, \quad 0 < i \leq k, \ N = 0, 1, 2, \ldots$$

*Proof.* We prove the Lemma by induction in $N$. Assume that $|u_N^i| < \gamma |u_N^0|$ for some $N$ and all $i > 0$. From the definition $\mathbf{u}_{N+1} = A(q^{N+1})^{-1} A(q^N) D(q^N) \mathbf{u}_N$. This implies the estimates
$$|u_{N+1}^i| < d^i(q^N)(|u_N^i| + C_1 \beta \delta_N \sum_{i=0}^{k} |u_N^i|), \quad \text{for } i > 0,$$

and

(3.3) $$|u_{N+1}^0| > d^0(q^N)(|u_N^0| - C_1 \beta \delta_N \sum_{i=0}^{n} |u_N^i|),$$

where $\delta_N = q^N - q^{N+1}$, and $C_1$ is some constant of the type described above.

Using the inductive assumption, and that $\delta_N < 1 - q$, we can rewrite these as
$$|u_{N+1}^i| < d^i(q^N)(|u_N^i| + C_2 \beta (1-q)|u_N^0|), \quad \text{for } i > 0,$$
$$|u_{N+1}^0| > d^0(q^N)(|u_N^0| - C_2 \beta (1-q)|u_N^0|).$$

Putting the two inequalities together, for $1 - q$ small we obtain
$$\left| \frac{u_{N+1}^i}{u_{N+1}^0} \right| < \frac{d_i(q^N)}{d^0(q^N)} \left( \left| \frac{u_N^i}{u_N^0} \right| + C_2 \beta (1-q) \right) (1 + 2C_2 \beta (1-q)).$$

Now, using the inductive hypothesis again, we can deduce
$$\left| \frac{u_{N+1}^i}{u_{N+1}^0} \right| < \alpha \gamma (1 + \frac{C_3}{\gamma} \beta (1-q)),$$

for some constant $C_3$. Now, let us choose the value of $C_{\rm r}$ in the Lemma to be equal to $C_3$. Then (3.2) implies that the right hand side of this inequality is less then $\gamma$. This completes the proof of the Lemma. $\square$

Now we can easily prove the second statement.

**Lemma 3.3.** *For $q$ sufficiently close to 1, for all $N \geq 0$ we have*

$$(3.4) \qquad |\log u^0_{N+1} - \sum_{n=0}^{N} \log d^0(q^N)| < C_{\rm s}\beta.$$

*Proof.* Indeed, since $B(x)$ is a compact family of matrices in $D_k$, we have $(1-\alpha)/\beta > 0$. This means that the conditions of Lemma 3.2 are satisfied for fixed small $\gamma$ if $q$ is close enough to 1. As in (3.3) we have the bounds

$$d^0(q^n)u^0_n(1 - C_4\beta\delta_n) < u^0_{n+1} < d^0(q^n)u^0_n(1 + C_4\beta\delta_n).$$

Taking logarithms and summing these inequalities we obtain

$$\sum_{n=0}^{N} \log d^0(q^n) + \log(1 - C_4\beta\delta_n) < \log u^0_{N+1} < \sum_{n=0}^{\infty} \log d^0(q^n) + \log(1 + C_4\beta\delta_n).$$

Finally, we can bound the error terms using that $|\log(1+x)| < 2|x|$ for small $|x|$, and that $\sum_{n=0}^{\infty} \delta_n = 1$. This implies the Lemma with the choice $C_{\rm s} = C_4$. $\square$

Note that clearly $\lim_{N\to\infty} \mathbf{e}^0(0)^t \mathbf{w}_N = \lim_{N\to\infty} u^0_N$. Also recall that we chose $\mathbf{w}_0$ close to the eigenvector $\mathbf{e}^0(1)$. Thus, multiplying both sides of (3.4) by $1-q$, and taking the limit $q \to 1$, we obtain

**Corollary 3.4.** *For any family $B : [0,1] \to D_k$, with the notation introduced above, we have*

$$\lim_{q \to 1}(1-q) \log \mathbf{e}^0(0)^t \ldots B(q^N) \ldots B(q^2)B(q)B(1)\mathbf{w}_0 = \lim_{q \to 1}(1-q) \log \prod_{N=0}^{\infty} d(q^N).$$

*Remark* 3.2. The conditions on $B(x)$ can be somewhat relaxed. The same proof works if the assumption of the $n$ different, positive, real eigenvalues is replaced by the following: there is a smooth family of non-singular matrices $A(x)$, which diagonalizes $B(x)$, such that $d^0(x)$, the 0th eigenvalue, is the eigenvalue of strictly maximal absolute value, and it is real, positive, and $\max_{x\in[0,1]} |d^i(x)|/d^0(x) < 1$ for $1 \leq i \leq k$. With these weakened conditions Corollary 3.4 can be applied to the case we study in our earlier paper [2]. $\square$

*Remark* 3.3. Corollary 3.4 cannot be directly applied to Proposition 3.1, since there our family $L_k(q,x)$ explicitly depends on $q$, and more importantly $L_k(1,0) \notin D_k$. These subtleties will be addressed in the Appendix. $\square$

## 4. Rational parametrization

In this Section we construct an explicit rational parametrization of the curve, defined by the characteristic equation of the matrix $L_k(1, x) = x^{-2k} M_k(x^2)^2$ and use it to evaluate the integral (3.1) as a sum of dilogarithms. To achieve this goal, we will find such a parametrization for the characteristic equation of the matrix $M_k(x^2)^{-1}$, separately for even and odd levels.

**4.1. Even level.** The matrix $M_k(x^2)$ acts on the $(k+1)$–dimensional linear space $B_k$, spanned by vectors $\mathbf{b}_i, i \in S_k$. There is an involution on $B_k$, which sends $\mathbf{b}_i$ to $x^{k-2i}\mathbf{b}_{k-i}$, and commutes with $M_k(x^2)$. The space $B_k$ splits into the direct sum of the subspaces $B_k^+$ and $B_k^-$ of invariants and anti-invariants with respect to this involution. Both of them are preserved by the action of $M_k(x^2)$.

The vectors $\mathbf{a}_i^+ = x^i \mathbf{b}_i + x^{k-i}\mathbf{b}_{k-i}, i = 0, 1, \ldots, n-2$, and $\mathbf{a}_n^+ = \mathbf{b}_{n-1}$, where $n = k/2+1$, form a linear basis in $B_k^+$. The vectors $\mathbf{a}_i^- = -x^i \mathbf{b}_i + x^{k-i}\mathbf{b}_{k-i}, i = 0, \ldots, n-2$, form a linear basis in $B_k^-$. In this new basis the matrix $M_k(x^2)$ breaks into two blocks: $M_k^+(x)$ of order $n$ and $M_k^-(x)$ of order $n-1$, respectively.

For $m > 1$ let us introduce the $m \times m$ matrices

$$N_m^+(x) = \begin{pmatrix} 1 & -x & 0 & \ldots & 0 & 0 & 0 \\ -x & 1+x^2 & -x & \ldots & 0 & 0 & 0 \\ 0 & -x & 1+x^2 & \ldots & 0 & 0 & 0 \\ \hdotsfor{7} \\ 0 & 0 & 0 & \ldots & 1+x^2 & -x & 0 \\ 0 & 0 & 0 & \ldots & -x & 1+x^2 & -x \\ 0 & 0 & 0 & \ldots & 0 & -2x & 1+x^2 \end{pmatrix}.$$

and

$$N_m^-(x) = \begin{pmatrix} 1 & -x & 0 & \ldots & 0 & 0 & 0 \\ -x & 1+x^2 & -x & \ldots & 0 & 0 & 0 \\ 0 & -x & 1+x^2 & \ldots & 0 & 0 & 0 \\ \hdotsfor{7} \\ 0 & 0 & 0 & \ldots & 1+x^2 & -x & 0 \\ 0 & 0 & 0 & \ldots & -x & 1+x^2 & -x \\ 0 & 0 & 0 & \ldots & 0 & -x & 1+x^2 \end{pmatrix}.$$

Let us also put $N_1^\pm(x) = 1$.

**Lemma 4.1.** *Let $n = k/2 + 1$.*
  (a) $N_n^+(x) = x^k(x^2 - 1)M_k^+(x)^{-1}$;
  (b) $N_{n-1}^-(x) = -x^k(x^2 - 1)M_k^-(x)^{-1}$;
  (c) *For $0 < x \leq 1$ the eigenvalues of the matrix $N_n^+(x)$ are different from the eigenvalues of the matrix $N_{n-1}^-(x)$.*

*Proof.* Parts (a) and (b) of the Lemma are proved in §6. For the proof of part (c) let us denote by $P_n(\mu, x)$ and $P_{n-1}^-(\mu, x)$ the characteristic polynomials of $N_n^+(x)$ and $N_{n-1}^-(x)$, respectively. We have:

$$P_n(\mu, x) = (1 + x^2 - \mu)P_{n-1}^-(\mu, x) - 2x^2 P_{n-2}^-(\mu, x).$$

Suppose that there is an eigenvalue $\xi$ of the matrix $N_n^+(x)$, which coincides with an eigenvalue of the matrix $N_{n-1}^-(x)$. Then $P_n(\xi, x) = P_{n-1}^-(\xi, x) = 0$, and so $P_{n-2}^-(\xi, x) = 0$. Since

$$(4.1) \qquad P_m^-(\mu, x) = (1 + x^2 - \mu)P_{m-1}^-(\mu, x) - x^2 P_{m-2}^-(\mu, x),$$

by induction we conclude that $P_1^-(\xi, x) = 0$. But $P_1^-(\mu, x) = 1$, a contradiction. □

Let $Q_n(\mu, x)$ be the characteristic polynomial of the matrix, which has the same entries as the matrix $N_n^+(x)$, except for the top left entry, which is replaced by $1+x^2$.

Clearly,

$$P_n(\mu, x) = Q_n(\mu, x) - x^2 Q_{n-1}(\mu, x).$$

On the other hand, the following recursion relations hold:

$$(4.2) \qquad Q_n(\mu, x) = (1 + x^2 - \mu)Q_{n-1}(\mu, x) - x^2 Q_{n-2}(\mu, x),$$

with the initial conditions: $Q_1 = 1 + x^2 - \mu, Q_0 = 2$.

Let us introduce new functions

$$(4.3) \qquad Z_n(\mu, x) = \frac{Q_n(\mu, x)}{(1 + x^2 - \mu)Q_{n-1}(\mu, x)}.$$

We obtain from (4.2):

$$(4.4) \qquad Z_n = 1 - \frac{x^2/(1 + x^2 - \mu)^2}{Z_{n-1}}.$$

The equation $P_n(\mu, x) = 0$ can be rewritten as

$$(4.5) \qquad Z_n(\mu, x) = \frac{x^2}{1 + x^2 - \mu}.$$

We would like to find an explicit parametrization for the solutions $x, \mu$ of the equation $P_n(\mu, x) = 0$.

Let us introduce a series of functions $f_j(t), j > 0$, satisfying the system of equations:

$$(4.6) \qquad (1 - f_1) = f_1 f_2^{-1}, \quad \text{and} \quad (1 - f_j)^2 = f_{j-1}^{-1} f_j^2 f_{j+1}^{-1}, \quad j > 0,$$

with $f_1(t) = t$.

Let us introduce the following rational functions in $t$: $x_n(t)^2 = f_n(t)/f_{n+1}(t)$, and $\mu_n(t) = f_n(t)$. Note that $P_n(\mu, x)$ is in fact a polynomial in $\mu$ and $x^2$, so that if we substitute these expressions, we will obtain a rational function in $t$.

**Proposition 4.2.** $P_n(\mu_n(t), x_n(t)) = 0$.

*Proof.* As was mentioned before, the equation $P_n = 0$ is equivalent to the equation (4.5). We will prove this equation by induction on $n$.

By definition, $Z_1 = Q_1/(1 + x^2 - \mu)Q_0 = 1/2$. On the other hand, we have: $x_1^2 = f_1/f_2 = 1 - t, \mu_1 = f_1 = t$, so $x_1^2/(1 + x_1^2 - \mu_1) = 1/2$. Hence the equation (4.5) is satisfied for $n = 1$.

Now let us assume that we have proved the Proposition for $n = m - 1$. By (4.5), it means that

$$Z_n(\mu_{m-1}, x_{m-1}) = \frac{x_{m-1}^2}{(1 + x_{m-1}^2 - \mu_{m-1})}. \tag{4.7}$$

Let us prove it for $n = m$.

We will use the fact that the expression

$$E_j(t) = \frac{x_j(t)^2}{(1 + x_j^2(t) - \mu_j(t))^2}$$

does not depend on $j$. Indeed, for $j > 1$,

$$E_j = \frac{f_j/f_{j+1}}{(1 + f_j/f_{j+1} - f_j)^2} = \frac{(1 - f_j)^2 f_{j-1}/f_j}{(1 + (1 - f_j)^2 f_{j-1}/f_j - f_j)^2}$$

$$= \frac{(1 - f_j)^2 f_{j-1}/f_j}{((1 - f_j)(1 + f_{j-1}/f_j + f_{j-1}))^2} = E_{j-1},$$

and therefore $E_j = E_{j-1} = \ldots = E_0 = \frac{1}{4(1-t)}$ for any $j \geq 0$.

This, together with the formula (4.4) and the initial condition $Z_1 = 1/2$, leads us to the conclusion that for each $m > 0$, $Z_m(\mu_j(t), x_j(t))$ does not depend on $j$.

*Remark* 4.1. In fact, $Z_m(\mu_j(t), x_j(t))$ can be expressed as a continued fraction, depending on $t$. This gives another way of finding rational parametrization for $\mu$ and $x$, which was used in [2]. □

In particular,

$$Z_m(\mu_m, x_m) = Z_m(\mu_{m-1}, x_{m-1}) = 1 - \frac{x_{m-1}^2/(1 + x_{m-1}^2 - \mu_{m-1})^2}{Z_{m-1}(\mu_{m-1}, x_{m-1})} = \frac{x_{m-1}^2 - \mu_{m-1}}{1 + x_{m-1}^2 - \mu_{m-1}},$$

where we used our inductive assumption (4.7) for $n = m - 1$. So, to prove formula (4.7) for $n = m$, we have to prove that

$$\frac{x_{m-1}^2 - \mu_{m-1}}{1 + x_{m-1}^2 - \mu_{m-1}} = \frac{x_m^2}{1 + x_m^2 - \mu_m}.$$

But the right hand side is equal to

$$\frac{(1-f_m)^2 f_{m-1}/f_m}{1+(1-f_m)^2 f_{m-1}/f_m - f_m} = \frac{(1-f_m)f_{m-1}/f_m}{1+f_{m-1}/f_m - f_{m-1}},$$

which is the same as the left hand side. □

It is convenient to pass to a new variable $y$, such that $t = -(y - y^{-1})^2/4$. One checks by induction that under this change of variables

$$(4.8) \qquad f_m = -\frac{(y-y^{-1})^2}{(y^{m-1}+y^{-m+1})^2}.$$

*Remark* 4.2. If $m$ is odd, then we can express $(y^{m-1} + y^{-m+1})/2$ as a polynomial $U_m(t)$ in $t$ of degree $(m-1)/2$. By the formula (4.8), in this case $f_m(t) = t/U_m^2(t)$. If $m$ is even, then $(y^{m-1} + y^{-m+1})(y + y^{-1})$ is expressible as a polynomial $U_m(t)$ in $t$ of degree $(m-2)/2$, and so $f_m(t) = t/[(1-t)U_m^2(t)]$. Note that the polynomials $U_m(t)$ are closely related to the Chebyshev polynomials. □

**Lemma 4.3.** *For any $0 < \alpha \leq 1$, there are $n$ solutions $t_1(\alpha) < \cdots < t_n(\alpha)$ to the equations $x_n(t) = \alpha$, all of which lie between $0$ and $1$. The numbers $\mu_n(t_1(\alpha)), \ldots, \mu_n(t_n(\alpha))$ – the eigenvalues of the matrix $N_n^+(\alpha)$ – are real and non-negative. The least among them is $\mu_n(t_1(\alpha))$.*

*Proof.* In our new variable $y$ the equation $x_n = \alpha$ can be rewritten as

$$\frac{y^n + y^{-n}}{y^{n-1} + y^{-n+1}} = \pm\alpha.$$

For $\alpha = 1$ we have the equation $(y^{2n-1} - 1)(y - 1) = 0$, whose solutions are $y = \pm e^{2\pi i l/(2n-1)}, l = 0, \ldots, 2n-2$. This gives $n$ solutions $t_p(1) = \sin^2 \frac{\pi(p-1)}{2n-1}, p = 1, \ldots, n$, to the equation $x_n(t) = 1$. For $\alpha = 0$ we have the equation $y^{2n} + 1 = 0$, so we conclude that the solutions to the equation $x_n(t) = 0$ are $t_{2j}(0) = \sin^2 \frac{\pi(2j-1)}{2n}, j = 1, \ldots, \frac{n}{2}$, each with multiplicity 2, if $n$ is even; and $t_{2j}(0) = \sin^2 \frac{\pi(2j-1)}{2n}, j = 1, \ldots, \frac{n-1}{2}$, with multiplicity 2, $t_n(0) = 1$ with multiplicity 1, if $n$ is odd.

We see that between two neighboring solutions $t_{2j-1}(1)$ and $t_{2j}(1), 1 \leq j \leq [\frac{n}{2}]$ of the equation $x_n(t) = 1$, there is always one solution $t_{2j}(0)$ of the equation $x_n(t) = 0$ of multiplicity 2. For odd $n$, there are two more solutions, $t_n(1)$ and $t_{n+1}(0) = 1$, of the respective equations. Therefore, for $\alpha$ between 0 and 1, there must be the following real solutions to the equation $x_n(t) = \alpha$: $t_{2j-1}(\alpha) \in [t_{2j-1}(1), t_{2j}(0)], t_{2j}(\alpha) \in [t_{2j}(0), t_{2j}(1)]$, for $1 \leq j \leq [\frac{n}{2}]$, and also a real solution $t_n(\alpha) \in [t_n(1), 1]$, if $n$ is odd. We have: $t_1(\alpha) < \cdots < t_n(\alpha)$.

Let us fix $x_n = \alpha$. As we have shown in the course of proving Proposition 4.2, for any $n$,
$$\frac{\alpha}{(1+\alpha^2 - \mu_n)^2} = \frac{1}{4(1-t)}.$$
Therefore,
(4.9) $$\mu_n = 1 + \alpha^2 \pm 2x_n\sqrt{1-t}.$$
Since $0 \leq t_p(\alpha) \leq 1$, hence $\mu_n(t_p(\alpha))$ is real and non-negative (in fact, $\mu_n(t) = 0$, if and only if $t = 0$). One can show that the sign is $+$ for the branches $t_p(\alpha)$ with $p$ even, and is $-$ for those with $p$ odd. Formula (4.9) shows that for $0 < \alpha \leq 1$, we have
$$\mu_n(t_1(\alpha)) < \mu_n(t_3(\alpha)) < \cdots < \mu_n(t_{n-1}(\alpha)) < 1 <$$
$$< \mu_n(t_n(\alpha)) < \cdots < \mu_n(t_4(\alpha)) < \mu_n(t_2(\alpha)),$$
for $n$ even, and the same for $n$ odd, with $\mu_n(t_{n-1}(\alpha))$ and $\mu_n(t_n(\alpha))$ switched. In particular, the least eigenvalue of the matrix $N^+(\alpha)$ is $\mu_n(t_1(\alpha))$. $\square$

Now we want to compare the eigenvalues of the matrices $N_n^+(x)$ and $N_{n-1}^-(x)$.

**Lemma 4.4.** *For any $0 < x \leq 1$ the eigenvalues of the matrix $N_{n-1}^-(x)$ are positive and real, and they are all greater then the least eigenvalue of the matrix $N_n^+(x)$.*

*Proof.* The matrix $N_{n-1}^-(x)$ is symmetric, therefore, all of its eigenvalues are real when $x$ is real. It follows from (4.1) that $\det(N_{n-1}^-(x)) = 1$ for all $x$ and $n > 1$. This implies that if we show the positivity of the eigenvalues of $N_{n-1}^-(x)$ for a particular $x \in \mathbb{R}$, then this property will hold for all $x \in \mathbb{R}$. Since $N^-(0)$ is the identity matrix, this property obviously holds for $x = 0$.

The proof of the other statement is similar. By the previous Lemma, the eigenvalues of $N_{n-1}^+(x)$ are all real and non-negative for $0 < x \leq 1$, and they do not coincide with the eigenvalues of $N_{n-1}^-(x)$ by Lemma 4.1 (c). Thus again, it is sufficient to check the statement for a particular $x$. Now we can take $x = 1$. The smallest eigenvalue of $N_{n-1}^+(1)$ is 0. It is clearly smaller than the eigenvalues of $N_{n-1}^-(1)$, which we just have shown to be positive. This completes the proof. $\square$

*Remark* 4.3. One can show that between two neighboring eigenvalues of the matrix $N_n^+(x)$ there is a unique eigenvalue of the matrix $N_{n-1}^-(x)$. $\square$

Recall that $N_n^+(x)$ and $-N_{n-1}^-(x)$ are the two blocks of the matrix $x^k(1-x^2)M_k(x^2)^{-1}$. Therefore, if $\mu$ is an eigenvalue of the matrix $N_n^+(x)$ or $N_{n-1}^-(x)$, then $\lambda = (1-x^2)^2\mu^{-2}$ is an eigenvalue of the matrix $L_k(1,x) = x^{-2k}M_k(x^2)^2$. Let us summarize our results.

**Proposition 4.5.** *For any $0 < \alpha \leq 1$, the matrix $L_k(1,\alpha)$ has $k+1$ different positive real eigenvalues, and the largest among them is $d_k(\alpha) = (1-\alpha^2)^2\mu_n(t_1(\alpha))^{-2}$, where $t_1(\alpha)$ is the unique solution to the equation $x_n(t) = \alpha$ from the interval $[0,\gamma]$, where $\gamma = \sin^2\frac{\pi}{2n}$.*

We will use this result to compute the integral (3.1).

According to Proposition 4.5,

$$\int_{x=0}^{x=1} \log d_k(x) \, d\log x = \int_{x=0}^{x=1} \log \sqrt{d_k(x)} \, d\log x^2 =$$

(4.10)
$$\int_{x=0}^{x=1} \log(1-x^2) \, d\log x^2 + \int_{t=0}^{t=\gamma} \log \mu_n(t) \, d\log x_n(t)^2$$

(note that $x(0) = 1$ and $x(\gamma) = 0$).

The first summand in formula (4.10) is equal to $-L(1)$. The computation of the second summand will be based on the inductive formula

(4.11)
$$\log f_m \, d\log \frac{f_m}{f_{m+1}} - \frac{1}{2} d\left[\log f_m \, \log \frac{f_m}{f_{m+1}}\right] =$$

$$2 \log f_m \, d\log(1 - f_m) - d\left[\log f_m \, \log(1 - f_m)\right] +$$

$$\log f_{m-1} \, d\log \frac{f_{m-1}}{f_m} - \frac{1}{2} d\left[\log f_{m-1} \, \log \frac{f_{m-1}}{f_m}\right],$$

which follows from (4.6) and the equality

$$\log f_m \, d\log \frac{f_{m-1}}{f_m} - \frac{1}{2} d\left[\log f_m \, d\log \frac{f_{m-1}}{f_m}\right] =$$

$$\log f_{m-1} \, d\log \frac{f_{m-1}}{f_m} - \frac{1}{2} d\left[\log f_{m-1} \, d\log \frac{f_{m-1}}{f_m}\right].$$

Recall that $\mu_n(t) = f_n(t)$, and $x_n(t)^2 = f_n(t)/f_{n+1}(t)$. From (4.11) we obtain by induction:

(4.12)
$$\int_{t=0}^{t=\gamma} \log \mu_n \, d\log x_n^2 = 2 \sum_{m=2}^{n} \int_{t=0}^{t=\gamma} \left(\log f_m \, d\log(1 - f_m) - \frac{1}{2} d[\log f_m \, \log(1 - f_m)]\right)$$

$$+ \int_{t=0}^{t=\gamma} \left(\log f_1 \, d\log \frac{f_1}{f_2} - \frac{1}{2} d\left[\log f_1 \, \log \frac{f_1}{f_2}\right]\right) + \int_{t=0}^{t=\gamma} \frac{1}{2} d[\log \mu_n \, \log x_n^2].$$

Since $\mu_n(0) = 1$ and $x_n(\gamma) = 1$, the last term is equal to 0. We know that $f_m(0) = 0$ for any $m$, therefore by the definition of the Rogers dilogarithm function $L(z)$,

(4.13)
$$\int_{t=0}^{t=\gamma} \log \mu_n \, d\log x_n^2 = 2 \sum_{m=2}^{n} L(\beta_m) + L(\beta_1),$$

where $\beta_m = f_m(\gamma) = \sin^2 \frac{\pi}{2n} / \cos^2 \frac{\pi(m-1)}{2n}$, by (4.8). In particular, $\beta_n = 1$.

Now put $\delta_j = \sin^2 \frac{\pi}{k+2} / \sin^2 \frac{\pi(j+1)}{k+2}, j = 1, \ldots, k$, where $k = 2n - 2$, as in formula (1.1). Clearly, $\delta_{(n-1)\pm(i-1)} = \beta_i, i = 1, \ldots, n-1$.

*Remark* 4.4. The formula (4.6) shows that the numbers $\delta_j$ satisfy the system of equations

$$(1 - \delta_j)^2 = \prod_{1 \leq i \leq k-1} \delta_i^{a_{ij}}, \quad 1 \leq j \leq k-1,$$

where $\|a_{ij}\|$ is the Cartan matrix of the type $A_{k-1}$ (this formula holds true for odd $k$ as well). This system appears in the asymptotic analysis of the thermodynamic Bethe Anzatz equation of the XXZ model [1]. $\square$

The right hand side of (4.13) coincides with

$$\sum_{j=1}^{k} L(\delta_j) + L(1).$$

So, we obtain from (4.10) and (4.13):

**Theorem 4.6.**

$$\int_{x=0}^{x=1} \log d_k(x)\, d\log x = \sum_{j=1}^{k} L(\delta_j).$$

This is equal to the left hand side of the identity (1.1) for $k = 2n$. Therefore, Theorem 4.6 together with Remark 3.1 and Theorem 2.3, proves the identity (1.1) for even values of $k$.

**4.2. Odd level.** Our strategy is again to decompose the matrix $M_k(x^2)$ into two blocks, and then study the eigenvalues of the inverse to them.

We use the same notations as at the beginning of the previous subsection. We have the subspaces $B_k^+$ and $B_k^-$ of the space $B_k$, generated by the vectors $\mathbf{a}_i^+, i = 0, \ldots, n-1$, and $\mathbf{a}_i^-, i = 0, \ldots, n-1$, where $n = (k+1)/2$. In this new basis the matrix $M_k(x^2)$ splits into two blocks $M_k^\pm(x)$ of order $n$.

Let us introduce the following $m \times m$ matrix

$$N_m^\pm(x) = \begin{pmatrix} 1 & -x & 0 & \ldots & 0 & 0 & 0 \\ -x & 1+x^2 & -x & \ldots & 0 & 0 & 0 \\ 0 & -x & 1+x^2 & \ldots & 0 & 0 & 0 \\ \multicolumn{7}{c}{\dotfill} \\ 0 & 0 & 0 & \ldots & 1+x^2 & -x & 0 \\ 0 & 0 & 0 & \ldots & -x & 1+x^2 & -x \\ 0 & 0 & 0 & \ldots & 0 & -x & 1 \mp x + x^2 \end{pmatrix}.$$

In the same way as in the previous subsection, we can prove

**Lemma 4.7.** (a) $N_n^\pm(x) = \pm x^k(x^2 - 1)M_k^\pm(x)^{-1}$.
(b) *For any $0 < x \leq 1$, the eigenvalues of the matrix $N_n^+(x)$ are different from the eigenvalues of the matrix $N_n^-(x)$.*

We denote by $R_n(\mu, x)$ the characteristic polynomial of the matrix $N_n^+(x)$. Let $S_n(\mu, x)$ be the characteristic polynomial of the matrix, which has the same entries as the matrix $N_n^+(x)$, except for the top left and the bottom right entry, which are replaced by $1 + x^2$. $R_n(\mu, x)$ can be rewritten as follows:

$$R_n(\mu, x) = S_n(\mu, x) - (x^2 + x)S_{n-1}(\mu, x) + x^3 S_{n-2}(\mu, x).$$

Using the recursion relation

$$S_n(\mu, x) = (1 + x^2 - \mu)S_{n-1}(\mu, x) - x^2 S_{n-2}(\mu, x),$$

we can rewrite the equation $R_n(\mu, x) = 0$ as

$$(1 - x - \mu)S_{n-1} = x^2(1 - x)S_{n-2}.$$

It is equivalent to the equation

$$(4.14) \qquad Y_{n-1}(\mu, x) = \frac{x^2(1-x)}{(1 + x^2 - \mu)(1 - x - \mu)},$$

where $Y_n(\mu, x)$ is defined by the formula

$$(4.15) \qquad Y_n(\mu, x) = \frac{S_n(\mu, x)}{(1 + x^2 - \mu)S_{n-1}(\mu, x)}.$$

It satisfies the recursion

$$(4.16) \qquad Y_n = 1 - \frac{x^2/(1 + x^2 - \mu)^2}{Y_{n-1}}.$$

Let us introduce functions $g_j(t), j > 0$, satisfying the system of equations

$$(4.17) \qquad (1 - g_1)^2 = g_1 g_2^{-1}, \quad \text{and} \quad (1 - g_j)^2 = g_{j-1}^{-1} g_j^2 g_{j+1}^{-1}, \quad j > 1,$$

with $g_1(t) = t$.

Introduce the following functions: $x_n(t) = \prod_{1 \le i \le n}(1 - g_i(t))$ (note that $x_n(t)^2 = g_n(t)/g_{n+1}$) and $\mu(t) = g_n(t)$.

**Proposition 4.8.**
$$R_n(\mu_n(t), x_n(t)) = 0.$$

*Proof.* We prove this statement by induction, following the main steps of the proof of Proposition 4.2. First we check this statement for $n = 2$. Then we show by induction that the expression

$$F_j(t) = \frac{x_j(t)}{1 + x_j(t)^2 - \mu_j(t)}$$

does not depend on $j$ and is equal to $1/(2 - t)$.

By our inductive assumption,

$$Y_{m-2}(\mu_{m-1}, x_{m-1}) = \frac{x_{m-1}^2(1 - x_{m-1})}{(1 + x_{m-1}^2 - \mu_{m-1})(1 - x_{m-1} - \mu_{m-1})}.$$

The formula (4.16) and the fact that $F_n$ does not depend on $n$ give us the formula

$$Y_{m-1}(\mu_m, x_m) = Y_{m-1}(\mu_{m-1}, x_{m-1}) = \frac{x_{m-1}(\mu_{m-1} + x_{m-1} - x_{m-1}^2)}{(1 + x_{m-1}^2 - \mu_{m-1})(1 - x_{m-1})}.$$

To prove the Proposition for $n = m$, we have to show that

$$\frac{x_{m-1}(\mu_{m-1} + x_{m-1} - x_{m-1}^2)}{(1 + x_{m-1}^2 - \mu_{m-1})(1 - x_{m-1})} = \frac{x_m^2(1 - x_m)}{(1 + x_m^2 - \mu_m)(1 - \mu_m - x_m)}.$$

Again using $F_m = F_{m-1}$, we reduce it to

$$\frac{\mu_{m-1} + x_{m-1} - x_{m-1}^2}{1 - x_{m-1}} = \frac{x_m - x_m^2}{1 - \mu_m - x_m}.$$

Let us subtract 1 from both sides of this formula. We obtain

$$\frac{2 - F_m^{-1}}{(1 - \mu_m)/x_m - 1} = \frac{2 - F_{m-1}^{-1}}{1/x_{m-1} - 1}.$$

But this formula follows from the formulas $1 - \mu_m = x_m/x_{m-1}$ and $F_m = F_{m-1}$. The Proposition is proved. $\square$

If we pass to a new variable $y$, such that $t = -(y - y^{-1})^2$, then our functions $g_m(t)$ will take form:

$$(4.18) \qquad g_m = -\frac{(y^2 - y^{-2})^2}{(y^{2m-1} + y^{-2m+1})^2}$$

which can be checked by induction.

*Remark* 4.5. We know that $(y^{2m-1} + y^{-2m+1})/(y + y^{-1})$ is a polynomial in $t$ of degree $m - 1$, which is equal to $U_{2m-1}(t/4)$, introduced in Remark 4.2. Therefore, $g_m(t) = t/U_{2m-1}^2(t/4)$. $\square$

Using formula (4.18), we can prove that the statement of Lemma 4.3 holds true in our case. Thus, for any $0 < \alpha \leq 1$ the least eigenvalue of the matrix $N_n(\alpha)$ is equal to $\mu_n(t_1(\alpha))$, where $t_1(\alpha)$ is the least solution of the equation $x_n(t) = \alpha$. It lies between 0 and $4\sin^2 \frac{\pi}{2(2n+1)}$.

Using Lemma 4.7 (b), we again prove that for $0 < x \leq 1$, all eigenvalues of the matrix $N_n^-(x)$ are greater than the least eigenvalue of the matrix $N_n(x)$. Lemma 4.7 (a) then implies that the statement of Proposition 4.5 holds true in this case, if we put $\gamma = 4\sin^2 \frac{\pi}{2(2n+1)}$.

We can therefore calculate the integral (3.1) in the same way as for the even level. The formulas (4.11) and (4.12) with $f_m$ replaced by $g_m$ hold true, and so we obtain

$$\int_{t=0}^{t=\gamma} \log \mu_n(t)\, d \log x_n^2(t) = 2 \sum_{m=1}^{n} L(\beta_m),$$

where now $\beta_m = g_m(\gamma) = \sin^2 \frac{\pi}{2n+1} / \cos^2 \frac{\pi(2m-1)}{2(2n+1)}$. The right hand side can be rewritten in terms of $\delta_j$ from the formula (1.1) for $k = 2n - 1$, as

$$\sum_{j=1}^{k} L(\delta_j) + L(1).$$

Formula (4.10) then gives as a result Theorem 4.6 for odd values of $k$. Together with Remark 3.1 and Theorem 2.3, this proves the identity (1.1) for odd $k$.

## 5. Connection with algebraic $K$-theory

In this Section we will discuss the connection between the dilogarithm identities and the group $K_3$ of totally real number fields. Namely, we will explain how to construct elements of torsion in $K_3$ of totally real number fields. Note that related questions have been recently discussed in [9, 10, 11].

One way to do that is through the so-called Bloch group, which is related to $K_3$ by Suslin's theorem. The Rogers dilogarithm function defines a homomorphism from the Bloch group to $\mathbb{R}/(\pi^2\mathbb{Z})$, and we will show that this homomorphism is injective on the torsion part of $K_3(\mathbb{R})$, using the identities (1.1).

The other way to construct elements of torsion in $K_3$ is to use its presentation, due to Bloch, through the relative group $K_2$ of the projective line modulo two points. This construction is close in spirit to the computation of integral in §4.

We also discuss the connections between these two approaches and with regulators.

**5.1. Bloch group.** Let us recall Suslin's theorem on the structure of the indecomposable part of the group $K_3(F)$, where $F$ is a field [12] (cf. also [13]). Let $D(F)$ be the group, generated over $\mathbb{Z}$ by formal symbols $[x], x \in F^\times \backslash \{1\}$, where $F^\times$ is the multiplicative group of $F$. Let $(F^\times \otimes F^\times)_a$ be the quotient of the group $F^\times \otimes_\mathbb{Z} F^\times$ by the subgroup, generated by elements $x \otimes y + y \otimes x$, and define a homomorphism $\phi : D(F) \to (F^\times \otimes F^\times)_a$ by the formula $\phi([x]) = x \otimes (1-x)$. The kernel $C(F)$ of this homomorphism contains the elements of the form

(5.1)
$$[x] - [y] + [y/x] - [(1-x^{-1})/(1-y^{-1})] + [(1-x)/(1-y)], \quad x \neq y \in F^\times \backslash \{1\}.$$

The quotient $B(F)$ of $C(F)$ by the subgroup, generated by the elements of this form, is called the Bloch group [14, 15].

Let $K_3^{\mathrm{ind}}(F)$ be the indecomposable part of $K_3(F)$, i.e. the cokernel of the product map $K_1(F)^{\otimes 3} \to K_3(F)$.

**Theorem 5.1** ([12], **Theorem 5.2**). *There is an exact sequence*

$$0 \to \mathrm{Tor}(F^\times, F^\times)\tilde{} \to K_3^{\mathrm{ind}}(F) \to B(F) \to 0,$$

*where* $\mathrm{Tor}(F^\times, F^\times)\tilde{}$ *is the unique non-trivial extension of* $\mathbb{Z}/2$ *by the group* $\mathrm{Tor}(F^\times, F^\times)$.

From now on let $F$ be a totally real field of algebraic numbers. Then $\mathrm{Tor}(F^\times, F^\times)\tilde{} \simeq \mathbb{Z}/4$. The element $[x] + [1-x]$ belongs to the Bloch group, does not depend on $x$, and is of order 6 (cf. [12], Proposition 1.1). It is known that $B(\mathbb{Q})$ is isomorphic to the cyclic group $\mathbb{Z}/6$, which this element generates. Thus, in particular, $K_3^{\mathrm{ind}}(\mathbb{Q}) = \mathbb{Z}/24$ [16].

Let $B'(F)$ be the quotient of $B(F)$ by the subgroup, generated by $[x] + [1-x]$. Then we obtain

**Corollary 5.2.** $B'(F) \simeq K_3^{\mathrm{ind}}(F)/K_3^{\mathrm{ind}}(\mathbb{Q})$.

One can use the Rogers dilogarithm function to define a map $\mathcal{L}' : B'(\mathbb{R}) \to \mathbb{R}/(\mathbb{Z}\frac{\pi^2}{6})$. Let us extend the function $L(x)$ from the interval $[0,1] \subset \mathbb{R}$ to the whole real axis as follows:

$x > 1$: $L(x) = \frac{\pi^2}{3} - L(1/x)$;
$x < 0$: $L(x) = -\frac{\pi^2}{6} + L(1/(1-x))$.

Let $\bar{\mathcal{L}}'$ be a map $D(\mathbb{R}) \to \mathbb{R}$, which sends $[x]$ to $L(x)$. We can restrict it to $C(\mathbb{R})$. It is well-known that if $\alpha$ is an elements of $C(\mathbb{R})$ of the form (5.1), then $\bar{\mathcal{L}}'(\alpha) = \frac{\pi^2}{6} \mod \pi^2$, and also $\bar{\mathcal{L}}'([x] + [1-x]) = \frac{\pi^2}{6}$. In fact, these relations define the Rogers dilogarithm uniquely in the class of continuous and piecewise differentiable functions (cf. [17, 18]).

We obtain a well-defined homomorphism of abelian groups $\mathcal{L}' : B'(\mathbb{R}) \to \mathbb{R}/(\mathbb{Z}\frac{\pi^2}{6})$.

Following [19, 20], we can introduce a slightly modified map $\bar{\mathcal{L}} = \bar{\mathcal{L}}' - \frac{\pi^2}{6} : D(\mathbb{R}) \to \mathbb{R}$. Then, if $\alpha$ is an element of $C(\mathbb{R})$ of the form (5.1), $\bar{\mathcal{L}}(\alpha) = 0 \mod \pi^2$. Hence this map gives rise to a well-defined homomorphism $\mathcal{L} : B(\mathbb{R}) \to \mathbb{R}/(\mathbb{Z}\pi^2)$.

The homomorphisms $\mathcal{L}$ and $\mathcal{L}'$ and can be used to study torsion in the Bloch group $B(F)$ (and therefore in $K_3^{\mathrm{ind}}(F)$) for totally real number fields, using the dilogarithm identities (1.1).

The connection between dilogarithm identities and torsion elements in $K_3^{\mathrm{ind}}(\mathbb{R})$ was suggested to us by Goncharov. It is also implicit in the works of Lichtenbaum [20] and Govindachar [10]. Let us also mention the recent preprint[11] by Dupont and Sah, who analize this question from the point of view of homology of discrete groups.

Let us explain how to construct torsion elements in $K_3^{\mathrm{ind}}(\mathbb{Q}(\zeta_l)^+)$, where $\zeta_l$ is a primitive $l$th root of unity, and $\mathbb{Q}(\zeta_l)^+$ is the real part of the cyclotomic field $\mathbb{Q}(\zeta_l)$, from the identities (1.1). Let us denote

$$\delta_j(l) = \sin^2 \frac{\pi}{l} / \sin^2 \frac{\pi(j+1)}{l}, \quad j = 1, \ldots, l-3, \quad l > 3.$$

Clearly, $\delta_j(l) \in \mathbb{Q}(\zeta_l)^+$. Let us introduce elements $\Delta_l \in D(\mathbb{Q}(\zeta_l)^+)$ as $\Delta_l = 2 \sum_{j=1}^{l-3} [\delta_j(l)]$.

**Lemma 5.3.** *For each $l > 3$, $\Delta_l \in C(\mathbb{Q}(\zeta_l)^+)$.*

*Proof.* We have to check that $\phi(\Delta_l) = 0$. According to Remark 4.4,

$$\phi(\Delta_l) = \sum_{j=1}^{l-3} \delta_j \otimes (1-\delta_j)^2 = \sum_{j=1}^{l-3} \sum_{i=1}^{l-3} a_{ij} \cdot \delta_j \otimes \delta_i = \sum_{i=1}^{l-3} 2\delta_i \otimes \delta_i - \sum_{i=1}^{l-4}(\delta_i \otimes \delta_{i+1} + \delta_{i+1} \otimes \delta_i) = 0.$$

□

*Remark* 5.1. In fact, a more careful analysis shows that if $l$ is odd, $\frac{1}{4}\Delta_l \in C(\mathbb{Q}(\zeta_l)^+)$, and if $l$ is even, but not divisible by 4, $\frac{1}{2}\Delta_l \in C(\mathbb{Q}(\zeta_l)^+)$. □

Thus the symbol $\Delta_l$ defines certain elements of $B(\mathbb{Q}(\zeta_l)^+)$ and $B'(\mathbb{Q}(\zeta_l)^+)$.

**Proposition 5.4.**　(a) *Let $F$ be a totally real number field, and $m_p$ – the maximal number $m \geq 0$, such that $F$ contains $\mathbb{Q}(\zeta_{p^m}^+)$. The symbols $\Delta_{p^{m_p}}$ generate the group $B(F) = K_3^{\mathrm{ind}}(F)/(\mathbb{Z}/4)$.*
　(b) *The symbol $\Delta_l$ generates the group $B'(\mathbb{Q}(\zeta_l)^+) = K_3^{\mathrm{ind}}(\mathbb{Q}(\zeta_l)^+)/K_3^{\mathrm{ind}}(\mathbb{Q})$.*

*Proof.* We will use the identity (1.1), and the description of the group $K_3^{\mathrm{ind}}$ of a totally real number field, which is due to Mercuriev and Suslin [22], and Levine [23].

According to this description, for a totally real number field $F$, the group $K_3^{\mathrm{ind}}(F)$ is isomorphic to the cyclic group of order $2 \prod p^{m_p}$, where the product is taken over all primes. For example, for $F = \mathbb{Q}$, $m_2 = 2, m_3 = 1$, and all other $m_p = 0$, therefore $K_3^{\mathrm{ind}}(\mathbb{Q}) = \mathbb{Z}/24$. Hence, the group $B(F)$ is cyclic of order $b(F) = \frac{1}{2} \prod p^{m_p}$.

On the other hand, by the identity (1.1),

$$\mathcal{L}(\Delta_l) = \left( \frac{2-l}{3}\pi^2 - \frac{2}{l}\pi^2 \right) \mod \pi^2.$$

In particular, $\mathcal{L}(\Delta_6) = \frac{\pi^2}{3} \mod \pi^2$. The symbol $\Delta_6 = 4[\frac{1}{3}] + 2[\frac{1}{4}] \in B(\mathbb{Q})$ belongs to $B(F)$. By subtracting, if needed, the symbol $(2-l)\Delta_6$ from $\Delta_{p^{m_p}}$, we obtain an element of $B(F)$, which the homomorphism $\mathcal{L} : B(F) \to \mathbb{R}/\mathbb{Z}\pi^2$ sends to an element of $\mathbb{R}/\mathbb{Z}\pi^2$ of the exact order $p^{m_p}$, if $p \neq 2$, and $2^{m_2-1}$, if $p = 2$. The order of $\Delta_{p^{m_p}}$ is at least that of its image, therefore these elements of $B(F)$ generate a cyclic group of order at least $b(F)$, hence they generate the whole group $B(F)$.

Now let $F$ be the field $\mathbb{Q}(\zeta_l)^+$. Then the order of $K_3^{\mathrm{ind}}$ is equal to $2\mathrm{l.c.m.}(12,l)$. Therefore the group $B'(\mathbb{Q}(\zeta_l)^+)$ is cyclic of order $l/[\mathrm{g.c.d.}(12,l)]$. On the other hand, $\Delta_l$ is an element of $B'(\mathbb{Q}(\zeta_l)^+)$, and according to the identity (1.1), $\mathcal{L}'(\Delta_l) =$ $-\frac{12}{l} \cdot \frac{\pi^2}{6}$ mod $\frac{\pi^2}{6}$. Thus, $\Delta_l$ generates a subgroup of $B'(\mathbb{Q}(\zeta_l)^+)$ of order at least $l/[\text{g.c.d.}(12, l)]$, and so it generates the whole group $B'(\mathbb{Q}(\zeta_l)^+)$. □

In the course of proving this Proposition, we have also showed that the images of the elements $\Delta_l$ under the homomorphisms $\mathcal{L}$ and $\mathcal{L}'$ have the same orders as the orders of these elements. This implies

**Corollary 5.5.** *For a totally real number field $F$ the homomorphisms $\mathcal{L} : B(F) \to \mathbb{R}/(\mathbb{Z}\pi^2)$ and $\mathcal{L}' : B'(F) \to \mathbb{R}/(\mathbb{Z}\frac{\pi^2}{6})$ are injective.*

It is known that the torsion subgroup of $B(\mathbb{R})$ is generated by the images of the groups $B(\mathbb{Q}(\zeta_l)^+)$ of real parts of cyclotomic fields, and is therefore isomorphic to $\mathbb{Q}/\mathbb{Z}$. Thus we see that it is generated by the symbols $\Delta_l$, and that the map $\mathcal{L}$ is injective on the torsion subgroup of $B(\mathbb{R})$. It is tempting to conjecture that this map is injective on the whole Bloch group $B(\mathbb{R})$. This would be equivalent to the following statement: if $\mathcal{L}(\alpha) = 0$ mod $\pi^2$, then $\alpha$ is a linear combination of five-term elements of the form (5.1). This conjecture is analogous to the Zagier conjecture [24] and the Milnor-Ramakrishnan conjecture [25] on linear independence of values of polylogarithms. We would like to thank Goncharov for pointing this out.

To make this analogy more precise, we would like to discuss the connection between the map $\mathcal{L}$ and the regulator map for $K_3^{\text{ind}}(\mathbb{R})$. The regulators are certain abstractly defined maps from algebraic $K$-groups to suitable cohomology, such as Beilinson-Deligne cohomology. In our case the regulator is a map $r : K_3^{\text{ind}}(\mathbb{C}) \to \mathbb{C}/\mathbb{Z}(2\pi i)^2$. It is an interesting problem to find explicit formulas for (at least some pieces) of such a map.

It is known that $K_3^{\text{ind}}(\mathbb{R})$ embeds into $K_3^{\text{ind}}(\mathbb{C})$. Under this embedding, the torsion part isomorphically maps to the torsion subgroup of $K_3^{\text{ind}}(\mathbb{C})$, which coincides with the subgroup $\text{Tor}(\mathbb{C}^\times, \mathbb{C}^\times)\tilde{\ }$ of $K_3^{\text{ind}}(\mathbb{C})$. Note that there is no torsion in the Bloch group $B(\mathbb{C})$. Because of that, we can not extend the map $\mathcal{L}$ from $B(\mathbb{R})$ to $B(\mathbb{C})$ (the five-term relation for the Rogers dilogarithm does not hold for complex arguments!).

But we can take the composition

$$K_3^{\text{ind}}(\mathbb{R}) \to K_3^{\text{ind}}(\mathbb{C}) \xrightarrow{r} \mathbb{C}/\mathbb{Z}(2\pi i)^2 \xrightarrow{\arg(\cdot)} \mathbb{R}/\mathbb{Z}(2\pi i)^2,$$

and then take the projection $\mathbb{R}/\mathbb{Z}(2\pi i)^2 \to \mathbb{R}/\mathbb{Z}\pi^2$ to kill the subgroup $\text{Tor}(\mathbb{R}^\times, \mathbb{R}^\times)\tilde{\ }$ of $K_3^{\text{ind}}(\mathbb{R})$. This will give us a map $B(\mathbb{R}) \to \mathbb{R}/\mathbb{Z}\pi^2$. We conjecture that this map coincides with the map $\mathcal{L}$.

*Remark* 5.2. Note that if one composes the regulator $r$ with the map $\text{Re}\log(\cdot) = \log|\cdot|$ instead of $\text{Im}\log(\cdot) = \arg(\cdot)$, then one obtains a homomorphism $K_3^{\text{ind}}(\mathbb{C}) \to \mathbb{R}$, which gives a map $B(\mathbb{C}) \to \mathbb{R}$, because $\text{Tor}(\mathbb{C}, \mathbb{C})\tilde{\ }$ is torsion. It is known that this map coincides with the famous Bloch-Wigner dilogarithm function [14]

$$D([x]) = -\text{Im}\int_0^x \log(1-z)\,d\log z + \arg(1-z)\log|z|.$$

□

If we are given a regulator map, say, $r : K_3^{\text{ind}}(\mathbb{C}) \to \mathbb{C}/\mathbb{Z}(2\pi i)^2$, and an embedding $\sigma : F \to \mathbb{C}$ of a number field $F$, we can take the composition $r_\sigma$ of $r$ with the corresponding embedding $K_3^{\text{ind}}(F) \to K_3^{\text{ind}}(\mathbb{C})$. Usually one studies the $\log |\cdot|$ part of the regulator $r$, which is called the Borel regulator. For instance, Borel's theorem states that the images of $K_3^{\text{ind}}(F)$, corresponding to all complex embeddings of $F$, form an integral lattice in the real linear space, whose volume is related to the value of the zeta-function of $F$ at 2 (cf., e.g., [21]).

Since we work with totally real number fields, the Borel regulator is trivial, but we can take instead the maps $K_3^{\text{ind}}(F) \to K_3^{\text{ind}}(\mathbb{R})$, corresponding to all real embeddings of our field and compose them with the map $\mathcal{L} : K_3^{\text{ind}}(\mathbb{R}) \to \mathbb{R}/(\mathbb{Z}/\pi^2)$ (which conjecturally coincides with the $\arg(\cdot)$ part of the regulator $r$). It turns out that the images of the symbols $\Delta_{k+2} \in B(\mathbb{Q}(\zeta_{k+2})^+)$ in $\mathbb{R}/(\mathbb{Z}\pi^2)$ under the composed maps $\mathcal{L}_\sigma$ are related to conformal dimensions of primary fields in the $\hat{\mathfrak{sl}}_2$ WZW theory of level $k$.

Indeed, different embeddings of the field $\mathbb{Q}(\zeta_{k+2})^+$ into $\mathbb{R}$ are labeled by integers $m = 1, \ldots, [\frac{k+2}{2}]$, which are relatively prime with $k+2$. Under the corresponding embedding $\sigma_m$, $\delta_j(k+2) = \sin^2 \frac{\pi}{k+2} / \sin^2 \frac{(j+1)\pi}{k+2} \in \mathbb{Q}(\zeta_{k+2})^+$ maps to $\delta_j^{(m)}(k+2) = \sin^2 \frac{\pi m}{k+2} / \sin^2 \frac{\pi m(j+1)}{k+2} \in \mathbb{R}$. Therefore the symbol $\Delta_{k+2} \in B(\mathbb{Q}(\zeta_{k+2})^+)$ maps to the symbol $\Delta_{k+2}^{(m)} = 2 \sum_{j=1}^{k-1} [\delta_j^m(k+2)] \in B(\mathbb{R})$.

We know that $\mathcal{L}_{\sigma_m}(\Delta_{k+2}) = \mathcal{L}(\Delta_{k+2}^{(m)})$ must be an element of the group $\mathbb{R}/(\mathbb{Z}\pi^2)$ of order $k+2$. The exact value was calculated by Kirillov in [26] (cf. formulas (3.1) and (3.7)), who proved a generalization of the identity (1.1). He showed that

$$\bar{\mathcal{L}}(\Delta_{k+2}^{(m)}) = \frac{1}{3}(c_k - 24 h_k^{(m-1)} + 6(m-1) - k)\pi^2, \tag{5.2}$$

where $c_k = 3k/(k+2)$ is the central charge and $h_k^{(m)} = (m^2 - 1)/4(k+2)$ is the conformal dimension of the primary field of spin $(m-1)/2$. From (5.2) we deduce

$$\mathcal{L}(\Delta_{k+2}^{(n)} - \Delta_{k+2}^{(m)}) = 8(h_k^{(m)} - h_k^{(n)})\pi^2 \mod \pi^2. \tag{5.3}$$

**5.2. Relative $K_2$.** Now we would like to discuss another construction of elements of the group $K_3^{\text{ind}}(F)$. This construction is motivated by the calculation of the integral (3.1) in §4, and was suggested to us by Bloch. It is based on the idea of Bloch that $K_3^{\text{ind}}(F)$ is connected with the relative $K$-group $K_2(\mathbb{P}_F^1; \{a, b\})$ of the projective line modulo two points $a, b \in \mathbb{P}_F^1$ [14].

Let $R$ be the semi-local ring of rational functions on $\mathbb{P}_F^1$, which are well-defined at the points $a$ and $b$. Let $J$ be its Jacobson ideal, which consists of functions vanishing at $a$ and $b$. There is the following exact sequence (cf., e.g., [23], p.328):

$$0 \to K_3^{\text{ind}}(F) \to K_2(R, J) \xrightarrow{i} K_2(R). \tag{5.4}$$

It is known [14, 27] that $K_2(R, J)$ is the quotient of the group $(1+J)^\times \otimes F(\mathbb{P}^1)^\times$ by the subgroup, generated by elements of the form $f \otimes (1-f)$. Here $(1+J)^\times$ stands for the multiplicative group of rational functions on $\mathbb{P}^1_F$, such that $f(a) = f(b) = 1$. Let us denote by symbol $\{f, g\}$ the image of $f \otimes g$ in $K_2(R, J)$. The group $K_2(R)$ has a similar presentation: it is the quotient of the group $R^\times \otimes R^\times$ by the subgroup, generated by elements of the form $\{f, 1-f\}$, where $R^\times$ denotes the multiplicative group of invertible elements of $R$. So, it is generated by symbols $\{f, g\}$, where both $f$ and $g$ do not vanish at $a$ and $b$. If $g$ is such a function, then the image of a symbol $\{f, g\} \in K_2(R, J)$ under the map $i$ is equal to $\{f, g\} \in K_2(R)$.

There is another exact sequence (cf. [14], p.54):

$$(5.5) \qquad 0 \to K_3^{\text{ind}}(F) \to K_2(R, J) \xrightarrow{\coprod_{x \in \mathbb{P}^1_F} T_x} \coprod_{x \in \mathbb{P}^1_F} F(x)^\times,$$

where $F(x)$ is the residue field at the point $x$, and $T_x$ is the tame symbol

$$T_x(\{f, g\}) = (-1)^{\text{ord}_x f \, \text{ord}_x g} \left( \frac{f^{\text{ord}_x g}}{g^{\text{ord}_x f}} \right)(x).$$

Recall that in §4 we introduced rational functions $f_n(t), n > 0$, and $g_n(t), n > 0$ (cf. formulas (4.6) and (4.17)). One easily checks that the solutions to the equation $\mu_n(t) = f_n(t) = 1$ coincide with the solutions to the equation $x_n(t) = 0$. Those were found in the course of proving Lemma 4.3: $t = u_{2r-1} = \sin^2 \frac{\pi(2r-1)}{2n}, r = 1, \ldots, n$. Likewise, the solutions to the equation $g_n(t) = 1$ are $t = v_{2r-1} = 4\sin^2 \frac{\pi(2r-1)}{2(2n+1)}, r = 1, \ldots, n$. Denote by $R^n_{p,q}$ the semi-local rings of rational functions on $\mathbb{P}^1_{\mathbb{Q}(\zeta_{2n})^+}$, which are well-defined at $u_p$ and $u_q$. Denote by $S^n_{p,q}$ the semi-local ring of rational functions on $\mathbb{P}^1_{\mathbb{Q}(\zeta_{2n+1})^+}$, which are well-defined at $v_p$ and $v_q$. We will also denote by $J$ the corresponding ideals. We will restrict ourselves with $p, q$, which are odd and relatively prime with $l = 2n$, or $2n+1$, respectively.

**Proposition 5.6.** *The symbols*

$$\Gamma^{(p,q)}_{2n} = \left\{ f_n(t)^2, \frac{f_{n-1}(t)}{f_n(t)} \right\} \quad \text{and} \quad \Gamma^{(p,q)}_{2n+1} = \left\{ g_n(t)^2, \frac{g_{n-1}(t)}{g_n(t)} \right\}$$

*belong to the kernels of the maps $K_2(R^n_{p,q}, J) \to K_2(R^n_{p,q})$ and $K_2(S^n_{p,q}, J) \to K_2(S^n_{p,q})$, and therefore define elements in $K_3^{\text{ind}}(\mathbb{Q}(\zeta_{2n})^+)$ and $K_3^{\text{ind}}(\mathbb{Q}(\zeta_{2n+1})^+)$, respectively.*

*Proof.* In the group $K_2(R^n_{p,q})$ we have:

$$\left\{ f_n, \frac{f_{n-1}}{f_n} \right\} = \left\{ f_{n-1}, \frac{f_{n-1}}{f_n} \right\} - \left\{ \frac{f_{n-1}}{f_n}, \frac{f_{n-1}}{f_n} \right\} =$$

$$2\{f_{n-1}, 1-f_{n-1}\} - \left\{\frac{f_{n-1}}{f_n}, \frac{f_{n-1}}{f_n}\right\} - \left\{\frac{f_{n-2}}{f_{n-1}}, \frac{f_{n-2}}{f_{n-1}}\right\} + \left\{f_{n-2}, \frac{f_{n-2}}{f_{n-1}}\right\},$$

besause $f_{n-1}/f_n = (1-f_{n-1})^2 f_{n-2}/f_{n-1}$, by (4.6). Continuing by induction, we obtain

$$\left\{f_n, \frac{f_{n-1}}{f_n}\right\} = 2\sum_{j=2}^{n-1}\{f_j, 1-f_j\} + \{f_1, 1-f_1\} - \sum_{j=1}^{n-1}\left\{\frac{f_j}{f_{j+1}}, \frac{f_j}{f_{j+1}}\right\}$$

(here we used the fact that $f_1/f_2 = 1 - f_1$). The last term is equal to

$$-\sum_{j=1}^{n-1}\left(\left\{\frac{f_j}{f_{j+1}}, 1-\frac{f_j}{f_{j+1}}\right\} + \left\{\frac{f_{j+1}}{f_j}, 1-\frac{f_{j+1}}{f_j}\right\}\right) - \left\{\frac{f_1}{f_n}, -1\right\}.$$

Therefore
$$\left\{f_n^2, \frac{f_{n-1}}{f_n}\right\} = 2\left\{f_n, \frac{f_{n-1}}{f_n}\right\} =$$

(5.6)
$$4\sum_{j=2}^{n-1}\{f_j, 1-f_j\} + 2\{f_1, 1-f_1\} + 2\sum_{j=1}^{n-1}\left(\left\{\frac{f_j}{f_{j+1}}, 1-\frac{f_j}{f_{j+1}}\right\} - \left\{\frac{f_{j+1}}{f_j}, 1-\frac{f_{j+1}}{f_j}\right\}\right),$$

and is therefore equal to 0 in $K_2(R_{p,q}^n)$ (we used the fact that for any function $f$, $2\{f, -1\} = \{f, 1\} = 0$). Thus, $\Gamma_{2n}^{(p,q)}$ lies in the kernel of the map $i$ of (5.4), and hence it defines an element of $K_3^{\text{ind}}(\mathbb{Q}(\zeta_{2n})^+)$.

The second case can be analized in the same way. The only difference is that instead of $\{f_1, 1-f_1\}$ we obtain $2\{g_1, 1-g_1\}$. □

*Remark* 5.3. (compare with Remark 5.1) More careful analysis shows that if $l$ is odd, $\frac{1}{4}\Gamma_l^{(p,q)} \in K_3^{\text{ind}}(\mathbb{Q}(\zeta_l)^+)$, and if $l$ is even but not divisible by 4, $\frac{1}{2}\Gamma_l^{(p,q)} \in K_3^{\text{ind}}(\mathbb{Q}(\zeta_l)^+)$. □

*Remark* 5.4. Note that in [22, 23] elements of torsion in $K_3^{\text{ind}}(\mathbb{C})$ were presented by symbols $X_{\omega,\zeta} = \{(1 + (\omega - 1)t)^l, \zeta\}$, where $\omega$ and $\zeta$ are $l$th roots of unity. These symbols belong to the kernel of the map $i$ from (5.4) (here we take $a = 0, b = 1$), and are clearly of $l$th order. They do not belong to $K_3^{\text{ind}}(F)$ of a totally real number field. The images of the symbols, constructed in Proposition 5.6, in $K_3^{\text{ind}}(\mathbb{C})$ should be equivalent to linear combinations of symbols $X_{\omega,\zeta}$. It would be interesting to construct this equivalence explicitly. □

Let us consider now the symbols

(5.7) $$4\sum_{j=2}^{n-1}([f_j(u_q)] - [f_j(u_p)]) + 2([f_1(u_q)] - [f_1(u_p)]) \in D(\mathbb{Q}(\zeta_{2n})^+),$$

(5.8) $$4\sum_{j=1}^{n-1}([g_j(v_q)] - [g_j(v_p)]) \in D(\mathbb{Q}(\zeta_{2n+1})^+).$$

Then the formula (5.6) shows that such elements are in the kernel of the map $\phi$ and therefore they define elements in the corresponding Bloch group $B(\mathbb{Q}(\zeta_l)^+)$. Indeed, formula (5.6) holds true, if we replace the symbols $\{f,g\}$ by the tensor products $f \otimes g$. We may then specialize this formula at $t = u_q$. It will give us a relation in the group $F^\times \otimes F^\times$, and therefore in $(F^\times \otimes F^\times)_a$. Since $f_n(u_q) = 1$, we obtain

$$0 = f_n^2(u_q) \otimes \frac{f_{n-1}(u_q)}{f_n(u_q)} = \phi(4 \sum_{j=2}^{n-1} [f_j(u_q)] + 2[f_1(u_q)])$$

plus sum of expressions of the form $\phi(2[z]+2[z^{-1}])$, which are equal to 0 in $(F^\times \otimes F^\times)_a$ and are moreover equivalent to 0 in $B(F)$ (can be written as a linear combination of 5-term elements (5.1)) for any $z \in F^\times \setminus \{1\}$. This shows that the symbol (5.7) lies in the Bloch group. In the same way we can show that the symbol (5.8) also lies in the Bloch group.

A simple computation, using the formulas (4.8) and (4.18), shows that the elements (5.7) and (5.8) of the Bloch groups $B(\mathbb{Q}(\zeta_l)^+)$ coincide with the elements $\Delta_l^{(q)} - \Delta_l^{(p)}$, constructed before. Moreover, in the same way as in §4, we can show that

$$-\mathrm{Re} \int_{u_p}^{u_q} \log \frac{f_{n-1}(t)}{f_n(t)} \, d\log f_n(t) = \mathcal{L}(\Delta_l^{(q)} - \Delta_l^{(p)}),$$

and therefore this integral is equal to (5.2).

*Remark* 5.5. One can extend this construction to an explicit map $\psi : \mathrm{Ker}(i) \to B(F)$, which, together with the map $\int : \mathrm{Ker}(i) \to \mathbb{R}/(\mathbb{Z}\pi^2)$, given by

$$\int(\{f,g\}) = -\mathrm{Re} \int_a^b \log g \, d\log f$$

(cf. [28, 29, 30]), fits into the commutative diagram:

$$\begin{array}{ccc} \mathrm{Ker}(i) & \xrightarrow{\psi} & B(F) \\ \int \downarrow & & \downarrow \mathcal{L} \\ \mathbb{R}/(\mathbb{Z}\pi^2) & = & \mathbb{R}/(\mathbb{Z}\pi^2). \end{array}$$

□

Unfortunately, we can not interpret our integral

$$\int_{t=0}^{t=\gamma} \log \mu(t) \, d\log x^2(t)$$

from §4 as the value of the map $\int$ on an element of the group $K_3^{\mathrm{ind}}(F)$, if we use the presentation of this group, described before. The reason is that while the group $K_2(R, J)$ is generated by symbols $\{f, g\}$, such that $f(a) = f(b) = 1$, our functions $x(t)$ and $\mu(t)$ have a different property. Namely, at the point $t = 0$ we have $x = 1$, and at the point $t = \gamma$ we have $x = 0$, but $\mu = 1$.

However, we can introduce a similar group $\tilde{K}_2(R, J)$. Consider the group $(1 + J_a) \otimes (1 + J_b)$, where $J_a$ is the ideal of $R$, which consists of functions, vanishing the point $a$. Define $\tilde{K}_2(R, J)$ as the quotient of this group by the subgroup, generated by elements of the form $f \otimes (1 - f)$. If $f, g \in R$ are two functions, such that $f(a) = 1$ and $g(b) = 1$, we denote by $\{f, g\}$ the image of $f \otimes g$ in $\tilde{K}_2(R, J)$. These symbols indeed generate a group. As the inverse to the symbol $\{f, g\}$, we can take $\{f^{-1}, g\}$, if $f(b) \neq 0$, and $\{f, g^{-1}(1-f)^{\text{ord}_a g}\}$, if $f(b) = 0$.

At the moment we do not have a direct construction, which would relate this group to $K_2(R, J)$. But we conjecture the analogue of the exact sequence (5.5) for a totally real number field $F$:

$$0 \to K_3^{\text{ind}}(\mathbb{Q}) \to K_3^{\text{ind}}(F) \to \tilde{K}_2(R, J) \overset{\coprod_{x \in \mathbb{P}_F^1} T_x}{\longrightarrow} \coprod_{x \in \mathbb{P}_F^1} F(x)^\times.$$

Denote by $\text{Ker}(T)$ the kernel of the tame symbol map. We expect that there is a commutative diagram

$$\begin{array}{ccc} \text{Ker}(T) & \overset{\psi'}{\longrightarrow} & B'(F) \\ \int\downarrow & & \downarrow \mathcal{L}' \\ \mathbb{R}/(\mathbb{Z}\frac{\pi^2}{6}) & = & \mathbb{R}/(\mathbb{Z}\frac{\pi^2}{6}). \end{array}$$

In particular, the element $\{\mu_n^2(t), x_n^2(t)\}$, corresponding to either even or odd level $l - 2$, belongs to $\text{Ker}(T)$ and it should map to the symbol $\Delta_l \in B'(\mathbb{Q}(\zeta_l)^+)$, where $l = 2n$, or $l = 2n+1$. Indeed, as we have shown in §4, $\int_0^\gamma \log \mu_n^2(t) \, d \log x(t)_n^2 = \mathcal{L}'(\Delta_l)$.

## 6. Generalization to higher rank Lie algebras

In this section we discuss the extension of our approach to the higher rank Lie algebras $\hat{\mathfrak{sl}}_n$, using the desription of the crystal basis, given in [3].

We begin with adapting the notation and results of §2 to this more general case. Let $V_n^k$ be the level $k$ vacuum representation of the affine Lie algebra $\hat{\mathfrak{sl}}_n$. Introduce the set

$$S_n^k = \{\mathbf{a} = (a_1, a_2, \ldots, a_n) \mid a_i \geq 0, a_1 + a_2 + \cdots + a_n = k\}.$$

Denote by $\tau : (a_1, a_2, \ldots, n) \mapsto (a_n, a_1, \ldots, a_{n-1})$ the cyclic permutation acting on $S_n^k$, and define the element $\mathbf{a}^k \in S_n^k$ by $\mathbf{a}^k = (k, 0, \ldots, 0)$. Now the basic path $\mu$ is the sequence $(\mathbf{a}^k, \tau \mathbf{a}^k, \tau^2 \mathbf{a}^k, \ldots)$, and the set of restricted paths $R_n^k$ is the set of sequences $\eta$ of elements of $S_n^k$, which coincide with $\mu$ except for a finite set of indices. The weight function $\omega$ is defined by the formula (2.1), where the function $H = H_n$ is given by $H_n(\mathbf{a}, \mathbf{b}) = \max(\mathbf{h}(\mathbf{a}, \mathbf{b}))$, where $\mathbf{h}(\mathbf{a}, \mathbf{b}) \in \mathbb{Z}^n$,

$$\tag{6.1} \mathbf{h}_i(\mathbf{a}, \mathbf{b}) = \sum_{j=1}^{i} a_j - \sum_{j=1}^{i-1} b_j,$$

and $\max : \mathbb{Z}^n \to \mathbb{Z}$ is the maximal component of a vector. Again, it follows from the results of [5] that

$$\operatorname{ch} V_n^k = \sum_{\eta \in R_n^k} q^{\omega(\eta)}.$$

*Remark* 6.1. Formula (6.1) for $H_n$ follows from the formula for $H_n$, given in [3], although this is not immediately obvious. □

Now for $\mathbf{a} \in S_n^k$ we can introduce the set

$$W_N^{\mathbf{a}} = \{\eta \in R_n^k \mid \eta_{nN-1} = \mathbf{a},\ \eta_j = \mu_j \text{ for } j \geq nN\},$$

its character $w_N^{\mathbf{a}} = \operatorname{ch} W_N^{\mathbf{a}}$ and the column vector $\mathbf{w}_N$ of these characters in a certain order, which is fixed once and for all. Then one can show the analog of Lemma 2.2:

$$\mathbf{w}_{N+1} = q^{-knN} M_k(q^{nN+n-1}) \ldots M_k(q^{nN+1}) M_k(q^{nN}) \mathbf{w}_N,$$

where $M_k(x)_{\mathbf{a},\mathbf{b}} = x^{H_n(\mathbf{a},\mathbf{b})}$, for $\mathbf{a}, \mathbf{b} \in S_n^k$.

Finally, we obtain an identity by comparing two expressions for the asymptotics of $\operatorname{ch} V_n^k$. On the one hand, from [8] we have

$$\tag{6.2} \lim_{q \to 1}(1-q) \log \operatorname{ch} V_n^k = \frac{\pi^2}{6} \frac{(n^2-1)k}{n+k}.$$

On the other hand, if one extends Proposition 3.1 to all $n$, then as in Remark 3.1 one obtains

$$\tag{6.3} \lim_{q \to 1}(1-q) \log \operatorname{ch} V_n^k = \int_{x=0}^{x=1} \log d(x)\, d\log x,$$

where $d(x)$ is the highest eigenvalue of $M_k(x)$.

It is challenging to find an expression for this integral as we have in Theorem 4.6 for $\hat{\mathfrak{sl}}_2$. It seems to be a much more complicated problem in general. Recall that the calculation in §4 was based on a formula for $M_k(x^2)^{-1}$ in a certain basis. (See Lemma 4.1 and Lemma 4.7.) The fact that this matrix was tri-diagonal allowed us to find an inductive formula for the characteristic polynomial $P(x, \lambda)$ of this matrix, which then led to a rational parametrization of the corresponding curve $P(x, \lambda) = 0$. Partial calculations we have made in the case of $\hat{\mathfrak{sl}}_3$ indicate that such a curve is not rational for higher rank Lie algebras. It is unclear how to calculate the integral in (6.3) in this case in terms of the dilogarithm function.

Note that by (6.2), this integral should be equal to the central charge of the level $k$ vacuum representation of $\hat{\mathfrak{sl}}_n$. Dilogarithm identities with such right hand side are known to exist [18], but for $n > 2$ they do not possess the $\mathbb{Z}_n$ symmetry, which is manifest in the structure of the matrices $M_k(x^n)$.

The rest of this section is devoted to the explanation of this symmetry and to the study of the matrices $M_k(x^n)$. Our main result will be a formula for $M_k(x^n)^{-1}$ for general $n$, which will show that for $n > 2$ this inverse matrix is also localized around the diagonal in a suitable sense. This will also provide a new understanding of the function $H_n$.

As a first step, we conjugate the matrix $M_k(x^n)$ by rescaling the basis vector corresponding to $\mathbf{a} \in S_n^k$ by $x^{r(\mathbf{a})}$, where $r : \mathbb{Z}^n \to \mathbb{Z}$ is given by $r(\mathbf{a}) = \sum_{i=1}^n i a_i$. The result is a matrix $\bar{M}$ with entries $\bar{M}_{\mathbf{a},\mathbf{b}} = x^{nH_n(\mathbf{a},\mathbf{b})+r(\mathbf{a})-r(\mathbf{b})}$. Using the fact that the sum of the components of both $\mathbf{a}$ and $\mathbf{b}$ is $k$, one finds that

$$n\mathbf{h}_i(\mathbf{a},\mathbf{b}) + r(\mathbf{a}) - r(\mathbf{b}) = k + r(\tau^{-i}(\mathbf{a})) - r(\tau^{-i+1}(\mathbf{b})).$$

In other words, we have

$$x^{-k}\bar{M}_{\mathbf{a},\mathbf{b}} = x^{\max_{1 \leq i \leq n} r(\tau^{-i}(\mathbf{a})) - r(\tau^{-i+1}(\mathbf{b}))}.$$

This formula shows that $\bar{M}$ commutes with the permutation matrix $P_\tau$ induced by $\tau$. It is convenient to introduce the matrix $\hat{M} = x^{-k} P_\tau \bar{M}$. Then $\hat{M}_{\mathbf{a},\mathbf{b}} = x^{\max(\bar{\mathbf{h}}(\mathbf{a}-\mathbf{b}))}$, where the component $\bar{\mathbf{h}}_i(\mathbf{c}) = r(\tau^{i-1}\mathbf{c})$. Note that the eigenvalues $\hat{M}$ differ from the eigenvalues of $M$ by certain $n$th roots of unity.

To write down a formula for $\hat{M}^{-1}$ we introduce a dual point of view. Let $\mathbf{z} = \{z_1, z_2, \ldots, z_n\}$ be a set of variables. Then an element $\mathbf{a}$ of $S_n^k$ can be viewed as the monomial $\mathbf{z}^{\mathbf{a}} = z_1^{a_1} \ldots z_n^{a_n}$. We can associate to the column of $\hat{M}$ corresponding to $\mathbf{b}$ the polynomial

$$P_{\mathbf{b}} = \sum_{\mathbf{a} \in S_n^k} \hat{M}_{\mathbf{a},\mathbf{b}} \mathbf{z}^{\mathbf{a}}.$$

Now we describe the inverse in these terms. Introduce the matrix $N$, which has rows corresponding to the polynomials $Q_{\mathbf{a}}$ via

$$Q_{\mathbf{a}} = \sum_{\mathbf{b} \in S_n^k} R_{\mathbf{a},\mathbf{b}} \mathbf{z}^{-\mathbf{b}},$$

where

$$Q_{\mathbf{a}} = \mathbf{z}^{-\mathbf{a}} \prod_{1 \leq i \leq n,\, a_i \neq 0} (1 - xz_i/z_{i-1}).$$

We have adopted the convention $z_0 = z_n$.

**Proposition 6.1.**

$$N\hat{M} = (1 - x^n)I,$$

where $I$ is the identity matrix. Equivalently, the constant coefficient in the expanded product $Q_{\mathbf{a}} P_{\mathbf{b}}$ equals $1 - x^n$ if $\mathbf{a} = \mathbf{b}$, and vanishes otherwise.

*Proof.* We will consider only the generic case, when all components of $\mathbf{a}$ are positive. The other cases can be treated in the same way.

We need to show that the coefficient of $\mathbf{z}^{\mathbf{a}}$ in the product

$$P_{\mathbf{b}} \prod_{i=1}^{n}(1 - xz_i/z_{i-1}) \tag{6.4}$$

is $1 - x^n$ if $\mathbf{a} = \mathbf{b}$ and vanishes otherwise.

Recall that $\hat{M}_{\mathbf{a},\mathbf{b}} = x^{\max(\bar{\mathbf{h}}(\mathbf{a}-\mathbf{b}))}$, where $\bar{\mathbf{h}}(\mathbf{a}-\mathbf{b})$ is the vector of the integers $r(\tau^i(\mathbf{a}-\mathbf{b}))$ for $i = 0, \ldots, n-1$. Define the set

$$D_{\mathbf{b}}^j = \{\mathbf{c} \in S_n^k \mid r(\tau^{j-1}(\mathbf{c} - \mathbf{b})) = \min_{1 \leq i \leq n} r(\tau^{i-1}(\mathbf{c} - \mathbf{b}))\}.$$

Without loss of generality we can assume that $\mathbf{a} \in D_{\mathbf{b}}^1$.

Let us write (6.4) as

$$((1 - xz_1/z_n)P_{\mathbf{b}}) \prod_{i=2}^{n}(1 - xz_i/z_{i-1}).$$

(1) The expanded polynomial $(1 - xz_1/z_n)P_{\mathbf{b}}$ has only one term with exponent from $D_{\mathbf{b}}^1$: this is the monomial $(1 - x^n)\mathbf{z}^{\mathbf{b}}$
(2) For $\mathbf{c} \notin D_{\mathbf{b}}^1$

$$\mathbf{z}^{\mathbf{c}} \prod_{i=2}^{n}(1 - xz_i/z_{i-1}) \tag{6.5}$$

has no terms with exponents from $D_{\mathbf{b}}$. If $\mathbf{c} = \mathbf{b}$, the above product has a single term in $D_{\mathbf{b}}^1$: the term $\mathbf{z}^{\mathbf{b}}$.

First note if $\mathbf{d} \in \mathbb{Z}^n$ is such that the sum of its components is 0, then $r(\tau^{-1}(\mathbf{d}) - \mathbf{d}) = nd_1$. This implies that the components of the vector $\mathbf{h}(\mathbf{c} - \mathbf{b})$ are all equal modulo $n$. Hence we can conclude that unless $\mathbf{c} = \mathbf{b}$,

$$r(\mathbf{c} - \mathbf{b}) + n \leq \max(\mathbf{h}(\mathbf{a} - \mathbf{b})). \tag{6.6}$$

Denote by $\mathbf{v}_i$ the vector in $\mathbb{Z}^n$, which has components $v_i^l$ equal to 1 if $l = i$, to $-1$ if $l = i - 1$, and 0 otherwise. By the definition of the function $r$ we have

$$r(\mathbf{v}_i) = \begin{cases} 1 - n & \text{if } i = 1, \\ 1 & \text{if } i \neq 1. \end{cases} \tag{6.7}$$

Now statement (1) can be rewritten as follows: for $\mathbf{c} \in D_{\mathbf{b}}^1$

$$\max(\mathbf{h}((\mathbf{c} - \mathbf{v}_1) - \mathbf{b})) + 1 = \max(\mathbf{h}(\mathbf{c} - \mathbf{b})),$$

unless $\mathbf{c} = \mathbf{b}$, in which case we have $\max(\mathbf{h}(-\mathbf{v}_1)) = n - 1$. Then combining (6.7) for $i = 1$ and (6.6) with the definition of $D_{\mathbf{b}}$ implies (1).

To prove statement (2), let $\mathbf{c} \notin D_{\mathbf{b}}^1$ and $i \neq 1$. Formula (6.7) implies that $r(\mathbf{c} + \mathbf{v}_i - \mathbf{b}) - r(\mathbf{c} - \mathbf{b}) = 1$ and also

$$\max_{1 \leq j \leq n} (r(\tau^{j-1}(\mathbf{c} + \mathbf{v}_i - \mathbf{b})) - r(\tau^{j-1}(\mathbf{c} - \mathbf{b}))) = 1.$$

We assumed that the first component of $\mathbf{h}(\mathbf{c} - \mathbf{b})$ is not minimal, and realized that adding $\mathbf{v}_i$ to $\mathbf{c}$ does not change this. Thus $\mathbf{c} + \mathbf{v}_i \notin D_{\mathbf{b}}^1$.

The terms, which appear in the product (6.5), all have exponents of the form $\mathbf{c} + \sum_{i \in T} \mathbf{v}_i$, where $T$ is a subset of the set $\{2, \ldots, n\}$. By induction, none of them belongs to $D_{\mathbf{b}}^1$. The case $\mathbf{c} = \mathbf{b}$ is treated the same way.

This proves statement (2) and the Proposition. □

**Corollary 6.2.** *The number of nonzero terms in $M^{-1}$ grows linearly with the size of $S_n^k$ as $k \to \infty$.*

*Remark* 6.2. This formula for the inverse of $\hat{M}$ is reminiscent of a Weyl-type character formula. It would be interesting to find an interpretation of the polynomial $P_{\mathbf{b}}$ as a deformed character, and of $Q$ as a deformed "Weyl denominator". □

The formula for $N = (1 - x^2)\hat{M}^{-1}$ is especially simple for $n = 2$. In this case, since

$$(1 - xz_1/z_2)(1 - xz_2/z_1) = (1 + x^2) - xz_1/z_2 - xz_2/z_1,$$

we obtain the $k \times k$ matrix

$$N = \begin{pmatrix} 1 & -x & 0 & \ldots & 0 & 0 & 0 \\ -x & 1+x^2 & -x & \ldots & 0 & 0 & 0 \\ 0 & -x & 1+x^2 & \ldots & 0 & 0 & 0 \\ \ldots\ldots\ldots\ldots\ldots\ldots\ldots\ldots\ldots\ldots\ldots\ldots\ldots\ldots\ldots\ldots \\ 0 & 0 & 0 & \ldots & 1+x^2 & -x & 0 \\ 0 & 0 & 0 & \ldots & -x & 1+x^2 & -x \\ 0 & 0 & 0 & \ldots & 0 & -x & 1 \end{pmatrix}.$$

By construction, the matrix $N$ commutes with the cyclic permutation matrix $P_\tau$. If we decompose this matrix with respect to the two characters of the group $\mathbb{Z}_2$, generated by $P_\tau$, we obtain the matrices $N^+$ and $N^-$ of §4. This proves Lemma 4.1 (a),(b), and Lemma 4.7 (a).

In general, $N$ can be decomposed according to the characters of the cyclic group $\mathbb{Z}_n$. We expect that the block, corresponding to the trivial character, will contain the least eigenvalue of the matrix $(1 - x^n)M_k(x^n)^{-1}$, and it is therefore of main interest according to the formula (6.3).

Finally, let us note that in [5] a combinatorial model of the crystal basis is constructed for almost all affine Kac-Moody algebras, in particular, for all untwisted affine algebras, for which the the corresponding finite dimensional simply-laced Lie group has non-trivial center. (This excludes $E_8^{(1)}$, $F_4^{(1)}$, and $G_2^{(1)}$.) The generalization of the results of this section to these cases should be straightforward.

## 7. Appendix

This section is a continuation of §3. We complete the proof of Proposition 3.1.

First note that the entries of $M_k(qx)M_k(x)$ are all divisible by $x^k$, thus $L_k(q,x)$ depends algebraically on $q$ and $x$. Our strategy will be to divide $[0,1]$ into two parts at $\varepsilon(q)$, where $\lim_{q\to 1}\varepsilon(q) = 0$. On the interval $[\varepsilon(q),1]$, we will approximate $L_k(q,x)$ by $L_k(1,x)$, and then we show that the "tail" of $L_k(q,x)$, when $x < \varepsilon(q)$ does not contribute to the asymptotics. We cannot simply approximate the matrix $L_k(q,x)$ by $L_k(1,x)$, because the matrix $L_k(1,0)$ is upper triangular with 1-s on the diagonal, thus the limit $L_k(1,q^N)\ldots L_k(1,q^2)L_k(1,q)\mathbf{w}$ as $N \to \infty$ does not exist in general.

Consider $L_k(q,x)$ for some fixed $q$, and $x$ such that $L_k(q,x') \in D_k$, when $x' \geq x$. Introduce the same objects for this family as the ones associated to $B(x)$ before Lemma 3.2. Thus let $A(q,x)$ be the matrix of eigenvectors of $L_k(q,x)$, $d^i(q,x)$ for its eigenvalues. We denote by $\mathbf{u}_N(q)$ the vector $\mathbf{w}_N(q)$ in the basis $A(q,x)$, etc. We will omit the explicit dependence on $q$ for $q = 1$.

We modify $\alpha$ and $\beta$ as follows:

$$\alpha_q(\varepsilon) = \max_{x\in[\varepsilon,1], 1<i\leq n} \frac{d^i(q,x)}{d^0(q,x)}, \quad \text{and} \quad \beta_q(\varepsilon) = \max_{x\in[\varepsilon,1]} \left|\frac{d}{dx}\frac{1}{\det(A(q,x))}\right|$$

Note that $\alpha_1(0) = 1$ and $\beta_1(0) = \infty$.

**Lemma 7.1.** *Fix a small $\gamma > 0$. Then there is a function $\varepsilon(q) > 0$ such that $\lim_{q\to 1}\varepsilon(q) = 0$, and*
  (1) $L_k(q,x) \in D_k$ for $\varepsilon(q) < x \leq 1$,
  (2) *the family $L_k(q,x)$, satisfies the inequality of Lemma 3.2 on the interval $[\varepsilon(q),1]$, i.e. there exists a constant $C_1$ independent of $q$, such that*

$$C_{\mathrm{r}}(1-q) < \gamma\frac{1-\alpha_q(\varepsilon(q))}{\beta_q(\varepsilon(q))}.$$

*Proof.* Proposition 4.5 of the previous section implies that $L_k(1,x) \in D_k$, for $0 < x \leq 1$. Thus (1) follows from the fact that $D_k$ is an open set and $L_k(q,x)$ is smooth. The second condition (2) can be also satisfied, since clearly $(1-\alpha_q(\varepsilon))/\beta_q(\varepsilon)$ is a continuous positive function of $q$ and $\varepsilon$ in the domain where $L_k(q,x) \in D_k$ for $\varepsilon < x \leq 1$. $\square$

Let $N_q$ be the largest $N$ for which the inequality $q^N > \varepsilon(q)$ holds.

**Lemma 7.2.** *The function $\varepsilon(q)$ can be chosen to approach 0 sufficiently slowly so that*

$$\lim_{q\to 1}(1-q)(\log w^0_{N_q}(q) - \sum_{N=0}^{\infty}\log d^0(q^N)) = 0$$

*Proof.* Note that, since the matrix $L_k(1,1)$ is of rank 1, for $q$ close to 1, one will have $\gamma|u_1^0(q)| > |u_1^i(q)|$, thus the initial condition in Lemma 3.2 will be satisfied. Then Lemma 7.1 combined with Lemma 3.2 and Lemma 3.3 give us the inequalities

$$\left|\frac{u_{N_q}^i(q)}{u_{N_q}^0(q)}\right| < \gamma, \quad 1 \leq i \leq k, \tag{7.1}$$

and

$$|\log u_{N_q}^0(q) - \sum_{N=0}^{N_q} \log d^0(q, q^N)| < C_s \beta_q(\varepsilon(q)),$$

On the one hand, since $d^0(q, x)$ is a continuous function of both $q$ and $x$, clearly one can choose $\varepsilon(q)$ in such a way that

$$(1-q)\left(\sum_{N=0}^{\infty} \log d^0(q^N) - \sum_{N=0}^{N_q} \log d^0(q, q^N)\right) = 0.$$

On the other hand, if $\gamma$ is chosen sufficiently small, say $\gamma < 1/(2k)$, then according to (7.1) $u_{N_q}^0$ will be the dominant component of the vector $\mathbf{w}_{N_q}(q)$ in the basis $A(q, q^{N_q})$. Also $w_{N_q}^0(q)$ is the dominant component of the same vector in the standard basis (see Lemma 7.3 below). Thus $|\log u_{N_q}^0 - \log w_{N_q}^0|$ is bounded uniformly in $q$. This proves the Lemma. □

It follows from Lemma 7.2 that Proposition 3.1 is proved if we show that the matrices $L_k(q, q^N)$ for $N > N_q$ do not contribute to the asymptotics. More precisely, it suffices to prove

$$\lim_{q \to 1}(1-q)(\log w_\infty^0(q) - \log w_{N_q}^0(q)) = 0. \tag{7.2}$$

The argument will be similar to the combination of Lemma 3.2 and Lemma 3.3 in §3. First we will estimate the ratio of the components of $\mathbf{w}_N$, then using this we will bound $w_N^0$.

**Lemma 7.3.** *There exists a constant $\rho > 0$ such that $w_N^i/w_N^{i-1} < \rho q^{2N/(k+1)}$, for $i = 1, 2, \ldots, k$.*

*Proof.* We prove the estimate by induction, and the $\rho$ will be chosen in the process of the proof. For $N = 0$, it holds trivially. Assume that it holds for $N$. Recall that by Lemma 2.2 we have the recursion

$$\mathbf{w}_{N+1} = y^{-k} M_k(qy) M(y) \mathbf{w}_N,$$

where $y = q^{2N}$ and $M_k(x) = \|x^{\max(i, k-j)}\|$, $i, j = 0, 1, \ldots, k$. To simplify our notation, let $r = \rho q^{2N/(k+1)} = \rho y^{\frac{1}{k+1}}$ and $L(y) = y^{-k} M_k(qy) M(y)$. In these terms, it is clearly sufficient to prove that $qrw_{N+1}^{i-1} > w_{N+1}^i$ for $1 \leq i \leq k$, since $q \leq q^{2/(k+1)}$. The key is the following obvious relation between the entries of $M_k(y)$ denoted by $m_{i,j}(y)$:

$$m_{i,j-1}(y) = ym_{i-1,j}(y), \quad i,j = 1,2,\ldots,k.$$

Hence we obtain

$$m_{i,j-1}(qy)m_{j-1,s}(y) = qm_{i-1,j}(qy)m_{j,s-1}, \quad i,j,s = 1,2\ldots,k.$$

Note that since $L(y) = M(qy)M(y)$, the entries of the matrix $L(y)$ are quadratic expressions in the entries of the matrix $M(y)$. Further, the equality above relates the summands of the entries $l_{i,s}(y)$ and $l_{i-1,s-1}(y)$ of the matrix $L(y)$. Taking into consideration the two remaining terms in these sums: $m_{i,k}(qy)m_{k,s}(y)/y^k$, which is greater than $y^i$, and $m_{i-1,0}(qy)m_{0,s-1}(y)/y^k$, which is greater than $0$, we can conclude that

$$(7.3) \qquad l_{i,s}(y) \leq ql_{i-1,s-1}(y) + y^i, \quad i,s = 1,2,\ldots,k.$$

Now we have

$$(7.4) \quad qrw_{N+1}^{i-1} - w_{N+1}^i = qr\sum_{s=1}^{k+1} l_{i-1,s-1}(y)w_N^{s-1} - \sum_{s=0}^{k} l_{i,s}(y)w_N^s$$

$$\geq qrl_{i-1,k}(y)w_N^k - l_{i,0}(y)w_N^0 + \sum_{s=1}^{k} ql_{i-1,s-1}(y)rw_N^{s-1} - l_{i,s}(y)w_N^s.$$

To simplify the notation, from now on we omit the index $N$, i.e. $\mathbf{w} = \mathbf{w}_N$. We need to show that this last expression is positive. First, we can reduce this statement to the inequality

$$(7.5) \qquad rw^k + \sum_{s=1}^{k} l_{i,s}(y)(rw^{s-1} - w^s) > (2k+1)y^iw^0.$$

using (7.3) and the trivial estimates ($q$ is close to 1)

$$(7.6) \qquad l_{i,0}(y) < (k+1)y^i \quad \text{and} \quad ql_{i-1,k}(y) > 1 \quad \text{for} \quad i = 1,\ldots,k.$$

Note that by our inductive assumption the $rw^{s-1} > w^s$, so the terms on the left hand side are all positive.

Now we separate two complementary cases:
1. There is an $s \geq 1$ such that $rw^{s-1} > 2w^s$.
In this case, let $s'$ be the smallest such $s$. Then we have

$$rw^{s'-1} > 2w^{s'} \quad \text{and} \quad w^{s'-1} \geq (r/2)^{s'-1}w_0.$$

Also note that as similarly to (7.6) we have $l_{i,s} \geq y^{\max(0,i-s)}/2$. Then the $s$th term of the sum in (7.5) can be bound from below by

$$\frac{1}{2}y^{i-1}\frac{1}{2}rw^{s'-1} > \frac{1}{2}y^{i-1}\left(\frac{r}{2}\right)^{s'}w^0.$$

This is clearly greater then the right hand side of (7.5) if $\rho$ is sufficiently large, say $\rho > 2^{k+3}k$.

2. The inequality $rw^{s-1} \leq 2w^s$ holds for all $s \geq 1$.
Then we can use the first term of (7.5). We have

$$rw^k \geq 2(r/2)^k w^0.$$

Again this is greater then the right hand side of (7.5) for an appropriately large $\rho$.

Thus in both cases we have shown that $qrw_{N+1}^{i-1} > w_{N+1}^i$. This completes the inductive step and the proof of the Lemma. □

Now we can easily prove (7.2). Since any entry of the matrix $L(q,x)$ is less than $k+1$, Lemma 7.3 gives us the bound

$$w_{N+1}^0 < (1 + k(k+1)\rho q^{2N/(k+1)})w_N^0.$$

Thus we have

$$(1-q)(\log w_\infty^0 - \log w_{N_q}^0) \leq (1-q)\sum_{N=N_q}^\infty \log(1 + k(k+1)\rho q^{2N/(k+1)}).$$

Using $\log(1+x) < x$ and denoting $k(k+1)\rho$ by $C$, we can bound this by

$$(1-q)C\varepsilon(q)^{2/(k+1)}\sum_{N=0}^\infty q^{2N/(k+1)} = C\varepsilon(q)^{2/(k+1)}\frac{1-q}{1-q^{2/(k+1)}} \leq \frac{1}{2}C(k+1)\varepsilon(q)^{2/(k+1)}.$$

As $\lim_{q\to 1}\varepsilon(q) = 0$, this bound implies (7.2), and the proof of Proposition 3.1 is now complete.

**Acknowledgements**. We are greatly indebted to S.Bloch, A.Goncharov, and M.Levine for explaining to us various questions of algebraic $K$-theory.

Department of Mathematics, Harvard University, Cambridge, MA 02138

Department of Mathematics, Massachusetts Institute of Technology, Cambridge, MA 02139